\documentclass
[showkeys,showpacs,preprintnumbers,byrevtex]{revtex4}
\usepackage{amsmath,amsfonts,amssymb,amscd,amsxtra,amsthm}
\usepackage{graphicx}
\usepackage{bm}
\begin{document}
\preprint{CYCU-HEP-10-01}
\title{$\Lambda(1520)$ photoproduciton off the proton target with Regge contributions}
\author{Seung-il Nam}
\email[E-mail: ]{sinam@cycu.edu.tw}
\affiliation{Department of Physics, Chung-Yuan Christian University, Chung-Li 32023, Taiwan}
\author{Chung-Wen Kao}
\email[E-mail: ]{cwkao@cycu.edu.tw}
\affiliation{Department of Physics, Chung-Yuan Christian University, Chung-Li 32023, Taiwan}
\date{\today}
\begin{abstract}
We investigate the $\Lambda(1520,3/2^{-})\equiv\Lambda^{*}$ photoproduction off the proton target beyond the resonance region within
a model including the Regge contributions, the tree-level diagrams with the nucleon and the
certain resonance intermediate states and one contact term.
The Reggeized propagators for the $K$ and $K^{*}$ exchanges in the $t$ channel are employed in a gauge-invariant manner.
We compute the angular and energy dependences of the cross section and some polarization observables, such as the photon-beam
asymmetry and the polarization-transfer coefficients. Our results qualitatively agree with the current experimental data.
We find that the Regge contributions are necessary to explain the high-energy data beyond $E_{\gamma}\approx4$ GeV, especially for the angular
dependences in the forward region. On the contrary, the polarization observables are insensitive to the Regge contributions due to the contact-term
dominance which is a consequence of gauge invariance.  We also calculate
the $K^{-}$-angle distribution function in the Gottfried-Jackson frame, using the polarization-transfer
coefficients in the $z$ direction. We find that it owns a complicated angle and energy dependences in the forward $K^{+}$ scattering region.
\end{abstract}
\maketitle
\section{Introduction}
The photoproduction of hyperons off the nucleon target, $\gamma N\to KY$ is important in hadron physics because it reveals the strangeness-related
interaction structures of hadrons. There have been abundant experimental and theoretical efforts dedicated to it. Many experimental collaborations,
such as CLAS at Jafferson laboratory~\cite{McNabb:2003nf,Bradford:2005pt,Bradford:2006ba}, LEPS at SPring-8
~\cite{Niiyama:2009zz,Kohri:2009xe,Hicks:2008yn,Muramatsu:2009zp}, etc, have performed energetic research activities for the $\Lambda$, $\Sigma$, and $\Xi$ photoproductions.
Up to the resonance region $\sqrt{s}\lesssim3$ GeV, a simple hadronic model including the tree-level diagrams with the nucleon and the
certain resonance intermediate states has successfully explained the experimental data~\cite{Janssen:2001pe,Nam:2005uq,Nam:2006cx,Oh:2007jd,Nam:2009cv}.
Although there are more complicated higher-order contributions such as the finite-state interactions~\cite{Chiang:2001pw} or hadronic loops~\cite{Ozaki:2007ka}, but one can reach the agreement between the model result and the data without those high-order contributions.

However, this simple model is hardly applied to high-energy regions, because it is only valid at relatively low energy. On the other hand, it is well known that various reactions at high energy and low momentum transfer have been well described by Regge theory. Therefore, to extend our model to higher energy without the sacrifice of the satisfactory description of the low
energy data, mesonic Regge trajectories, corresponding to all the
meson exchanges with the same quantum numbers but different spins in
the $t$ channel at tree level, were employed
~\cite{Corthals:2005ce,Corthals:2006nz,Ozaki:2009wp}. This Regge
description is supposed to be valid in the limit
$(s,|t|\,\mathrm{or}\,|u|)\to(\infty,0)$~\cite{Regge:1959mz}. Even
in the resonance region, it has been argued that the Regge
description is still applicable to a certain
extent~\cite{Ozaki:2009wp,Sibirtsev:2003qh}.

In this article, we investigate the $\Lambda(1520,3/2^{-})\equiv\Lambda^{*}$ photoproduction off the proton target, $\gamma p\to K^{+}\Lambda^{*}$,
beyond the resonance region with an extended model including the original hadronic model and the interpolated Regge contributions.
This reaction has been intensively studied by CLAS and LEPS collaborations recently.
As shown in Refs.~\cite{Nam:2005uq}, up to the resonance region, this production process is largely dominated by the contact-term contribution.
This interesting feature supported by the experiments~\cite{Nakano,Muramatsu:2009zp} is a consequence of gauge invariance in a certain
description for spin-$3/2$ fermions, i.e. the Rarita-Schwinger formalism~\cite{Rarita:1941mf,Nath:1971wp}. The contact-term dominance simplifies
the analyses of the production process to a great extent. For instance, according to it, 1) one can expect a significant difference in the production
strengths between the proton- and neutron-target experiments as long as the coupling strength $g_{K^{*}N\Lambda^{*}}$ is small~\cite{Nam:2005uq},
and 2) the computation of the polarization-transfer coefficients is almost without any unknown parameter~\cite{Nam:2009cv}.

To date, there have been only a few experimental data of the $\Lambda^{*}$ photoproduction off the proton target~
\cite{Boyarski:1970yc,Barber:1980zv,Muramatsu:2009zp,Kohri:2009xe}. Among them, Barber et al. for LAMP2 collaboration explored
the process up to $E_{\gamma}\approx4.8$ GeV~\cite{Barber:1980zv}, whereas Muramatsu et al. for LEPS collaboration did it up to $2.4$
GeV~\cite{Muramatsu:2009zp}. Both of them were focusing on the resonance region below $\sqrt{s}\lesssim3$ GeV.
Hence, as done in Ref.~\cite{Nam:2005uq}, a simple model with the Born diagrams accompanied with the contact-term
dominance can reproduce the data qualitatively well. However, at $E_{\gamma}=11$ GeV as done by Boyarski et al.~\cite{Boyarski:1970yc},
this simple model fails in the forward region. Nevertheless it still agrees with the data qualitatively well
beyond $|t|\approx0.2\,\mathrm{GeV}^{2}$. To solve this discrepancy in
the high-energy forward region, we are motivated to introduce the Regge description which contributes significantly in the limit $(s,|t|)\to(\infty,0)$
by construction. Assuming that the Regge contributions still remain non negligible in the limit $(s,|t|)\to(s_{\mathrm{threshold}},\mathrm{finite})$,
we introduce an interpolating ansatz between the two ends.
After fixing the parameters including a cutoff mass for the form factors, it is straightforward to calculate all the physical observables.
We then present the ($\sigma$ and $d\sigma/d\Omega$ as a function of $E_{\gamma}$) and angular dependences ($d\sigma/dt$ and $d\sigma/d\Omega$
as a function of $\theta$), photon-beam asymmetry ($\Sigma$), and polarization-transfer coefficients ($C_{x,1/2}$, $C_{x,3/2}$, $C_{z,1/2}$,
and $C_{z,3/2}$) of the production process. Here, $\theta$ stands for the angle between the incident photon and outgoing kaon in the center of mass frame. Furthermore the $K^{-}$-angle distributions function, $\mathcal{F}_{K^{-}}$ in the Gottfried-Jackson frame using $C_{z,1/2}$ and $C_{z,3/2}$
are also calculated.
This article is organized as follows: In Section II, we provide the general formalism to compute the $\Lambda^{*}$ photoproduciton.
Numerical results are given in Section III. Finally, Section IV is for summary and conclusion.

\section{General formalism}
\subsection{Feynman amplitudes}
\begin{figure}[ht]
\includegraphics[width=10cm]{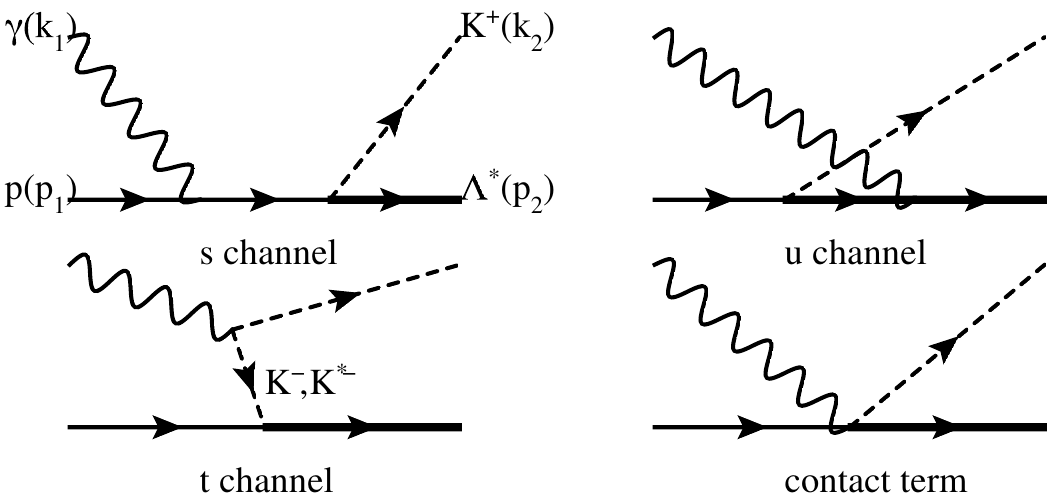}
\caption{Relevant Feynman diagrams for the $\Lambda(1520)$ photoproduction off the nucleon target.}
\label{FIG1}
\end{figure}

The general formalism of the $\gamma(k_1)+ N(p_1)\to K(k_2)+\Lambda^{*}(p_2)$ reaction process is detailed here.
The relevant Feynman diagrams are shown in Fig.~\ref{FIG1}. Here, we include the nucleon-pole and nucleon-resonance contributions in
the $s$-channel, $\Lambda^{*}$-pole contribution in the $u$-channel, $K$- and $K^{*}$-exchanges in the $t$-channel, and the contact-term contribution.
The relevant interactions are given as follows:
\begin{eqnarray}
\label{eq:GROUND}
{\cal L}_{\gamma KK}&=&
ie_K\left[(\partial^{\mu}K^{\dagger})K-(\partial^{\mu}K)K^{\dagger}
\right]A_{\mu},
\nonumber\\
{\cal L}_{\gamma
NN}&=&-\bar{N}\left[e_N\rlap{/}{A}-\frac{e_Q\kappa_N}{4M_N}\sigma\cdot
F\right]N,
\cr \mathcal{L}_{\gamma\Lambda^*\Lambda^*}&=&
-\bar{\Lambda}^{*\mu}
\left[\left(-F_{1}\rlap{/}{\epsilon}g_{\mu\nu}+F_3\rlap{/}{\epsilon}\frac{k_{1
\mu}k_{1
\nu}}{2M^{2}_{\Lambda^*}}\right)-\frac{\rlap{/}{k}_{1}\rlap{/}{\epsilon}}
{2M_{\Lambda^*}}\left(-F_{2}g_{\mu\nu}+F_4\frac{k_{1\mu}k_{1 \nu}}
{2M^{2}_{\Lambda^*}}\right)\right] \Lambda^{*\nu}+\mathrm{h.c.}, \cr
\mathcal{L}_{\gamma KK^{*}}&=& g_{\gamma
KK^{*}}\epsilon_{\mu\nu\sigma\rho}(\partial^{\mu}A^{\nu})
(\partial^{\sigma}K)K^{*\rho}+\mathrm{h.c.} \cr
\mathcal{L}_{KN\Lambda^*}&=&\frac{g_{KN\Lambda^*}}{M_{\Lambda^*}}
\bar{\Lambda}^{*\mu}\partial_{\mu}K\gamma_5N\,+{\rm h.c.}, \cr
\mathcal{L}_{K^{*}N\Lambda^*}&=&
-\frac{iG_{1}}{M_{V}}\bar{\Lambda}^{*\mu}\gamma^{\nu}G_{\mu\nu}N
-\frac{G_{2}}{M^{2}_{V}}\bar{\Lambda}^{*\mu}
G_{\mu\nu}\partial^{\nu}N
+\frac{G_{3}}{M^{2}_{V}}\bar{\Lambda}^{*\mu}\partial^{\nu}G_{\mu\nu}N+\mathrm{h.c.},
\cr {\cal L}_{\gamma
KN\Lambda^*}&=&-\frac{ie_Ng_{KN\Lambda^*}}{M_{\Lambda^*}}
\bar{\Lambda}^{*\mu} A_{\mu}K\gamma_5N+{\rm h.c.},
\end{eqnarray}
where $e_{h}$ and $e_{Q}$ stand for the electric charge of a hadron $h$ and unit electric charge, respectively. $A$, $K$, $K^{*}$, $N$, and
$\Lambda^{*}$ are the fields for the photon, kaon, vector kaon, nucleon, and $\Lambda^{*}$. As for the spin-$3/2$ fermion field, we use the Rarita-Schwinger (RS)
vector-spinor field~\cite{Rarita:1941mf,Nath:1971wp}. We use the notation $\sigma\cdot F=\sigma_{\mu\nu}F^{\mu\nu}$, where $\sigma_{\mu\nu}=i(\gamma_{\mu}\gamma
_{\nu}-\gamma_{\nu}\gamma_{\mu})/2$ and the EM field strength tensor $F^{\mu\nu}=\partial^{\mu}A^{\nu}-\partial^{\nu}A^{\mu}$. $\kappa_{N,\Lambda^{*}}$
denotes the anomalous magnetic moments for $N$ and $\Lambda^{*}$. Although the spin-$3/2$ $\Lambda^{*}$ has four different electromagnetic form factors
$F_{1,2,3,4}$ as shown in Eq.~(\ref{eq:GROUND}), we only take into account the dipole one ($F_{2}\equiv e_{Q}\kappa_{\Lambda^{*}}$) but ignore the monopole
($F_{1}\equiv e_{\Lambda^{*}}=0$), the quadrupole ($F_{3}$), and the octupole ($F_{4}$) ones, since their contributions are negligible~\cite{Gourdin:1963ub}.
Using the $\gamma KK^{*}$ interaction given in Eq.~(\ref{eq:GROUND}) and experimental data~\cite{Amsler:2008zzb},
one can easily find that $g_{\gamma K^{*\pm}K^{\mp}}=0.254/\mathrm{GeV}$. The strength of the $g_{KN\Lambda^{*}}$ can be extracted from the experimental data of
the full and partial decay widths: $\Gamma_{\Lambda^{*}}\approx15.6$ MeV and $\Gamma_{\Lambda^{*}\to\bar{K}N}/\Gamma_{\Lambda^{*}}\approx0.45$~\cite{Amsler:2008zzb}.
The decay amplitude for $\Lambda^{*}\to \bar{K}N$ reads:
\begin{equation}
\label{eq:decay}
\Gamma_{\Lambda^{*}\to\bar{K}N}=\frac{g^{2}_{KN\Lambda^*}
|{\bm p}_{\bar{K}N}|}
{4\pi M^{2}_{\Lambda^{*}}M^{2}_{K}}
\left(\frac{1}{4}\sum_{\mathrm{spin}}|
\mathcal{M}_{\Lambda^{*}\to\bar{K}N}|^{2}\right)
,\,\,\,\,\,i\mathcal{M}_{\Lambda^{*}\to\bar{K}N}
=\bar{u}(q_{\Lambda^{*}})\gamma_{5}q_{\bar{K}}^{\mu}u_{\mu}(q_{N}),
\end{equation}
where ${\bm p}_{\bar{K}N}$ indicates the three momentum of the decaying particle which can be obtained by the K\"allen function for a decay $1\to2,3$~\cite{Amsler:2008zzb}:
\begin{equation}
\label{eq:kollen}
{\bm p}_{23}=\frac{\sqrt{[M^{2}_{1}-(M_{2}+M_{3})^{2}][M^{2}_{1}-(M_{2}-M_{3})^{2}]}}{2M_{1}}.
\end{equation}
Here $M_{i}$ stands for the mass of the $i$-th particle. Substituting the experimental information into Eq.~(\ref{eq:decay}) and using Eq.~(\ref{eq:kollen}),
one is led to $g_{KN\Lambda^*}\approx11$. As for the $K^{*}N\Lambda^{*}$ interaction, there are three individual terms~\cite{Oh:2007jd}, and we
defined a notation $G_{\mu\nu}=\partial_{\mu}K^{*}_{\nu}-\partial_{\nu}K^{*}_{\mu}$. Since we have only insufficient experimental and
theoretical~\cite{Hyodo:2006uw} information to determine all the coupling strengths for $G_{1,2,3}$,
we set $G_{2}$ and $G_{3}$ to be zero for simplicity. The scattering amplitudes for the reaction processes have be evaluated
as follows:
\begin{eqnarray}
\label{eq:AMP}
i\mathcal{M}_{s}&=&-\frac{g_{KN\Lambda^*}}{M_{K}}
\bar{u}^{\mu}_2k_{2\mu}{\gamma}_{5}
\left[\frac{e_{N}[(\rlap{/}{p}_{1}+M_{N})F_{c}+\rlap{/}{k}_{1}F_{s}]}
{s-M^{2}_{N}}\rlap{/}{\epsilon}
-\frac{e_{Q}\kappa_{p}}{2M_{N}}
\frac{(\rlap{/}{k}_{1}+\rlap{/}{p}_{1}+M_{N})F_{s}}
{s-M^{2}_{N}}
\rlap{/}{\epsilon}\rlap{/}{k}_{1}\right]u_1,
\cr
i\mathcal{M}_{u}&=&-\frac{e_{Q}g_{KN\lambda}\kappa_{\Lambda^*}F_{u}}
{2M_{K}M_{\Lambda}}
\bar{u}^{\mu}_{2}\rlap{/}{k}_{1}\rlap{/}{\epsilon}
\left[\frac{(\rlap{/}{p}_{2}-\rlap{/}{k}_{1}+M_{\Lambda^{*}})}
{u-M^{2}_{\Lambda^{*}}}
 \right]
k_{2\mu}\gamma_{5}u_{1},
\cr
i\mathcal{M}^{K}_{t}&=&\frac{2e_{K}g_{KN\Lambda^*}F_{c}}{M_K}
\bar{u}^{\mu}_2
\left[\frac{(k_{1\mu}-k_{2\mu})(k_{2}\cdot\epsilon)}
{t-M^{2}_{K}} \right]
\gamma_{5}u_1,
\cr
i\mathcal{M}^{K^{*}}_{t}&=&
-\frac{ig_{\gamma{K}K^*}g_{K^{*}NB}F_{v}}{M_{K^{*}}}
\bar{u}^{\mu}_{2}\gamma_{\nu}
\left[\frac{(k^{\mu}_{1}-k^{\mu}_{2})g^{\nu\sigma}-
(k^{\nu}_{1}-k^{\nu}_{2})g^{\mu\sigma}
}{t-M^{2}_{K^*}}\right]
(\epsilon_{\rho\eta\xi\sigma}k^{\rho}_{1}
\epsilon^{\eta}k^{\xi}_{2})u_1,
\cr
i\mathcal{M}_{\mathrm{cont.}}
&=&\frac{e_{K}g_{KN\Lambda^*}F_{c}}{M_K}
\bar{u}^{\mu}_2\epsilon_{\mu}{\gamma}_{5}u_1,
\label{amplitudes}
\end{eqnarray}
where $s$, $t$, and $u$ indicate the Mandelstam variables, while $\epsilon$, $u_{1}$, and ${u}^{\mu}_{2}$
denote the photon polarization vector, nucleon spinor, and RS vector-spinor, respectively.

Since hadrons are not point-like, it is necessary to introduce the form factors representing their spatial distributions. It is rather technical to
include the form factors at the same time to preserve gauge invariance of the invariant amplitude. For this purpose, we employ the scheme developed in
Refs.~\cite{Haberzettl:1998eq,Davidson:2001rk}. This scheme preserves not only the Lorentz invariance but also the crossing symmetry of the invariant amplitude,
on top of the gauge invariance. Moreover, it satisfies the on-shell condition for the form factors: $F(q^{2}=0)=1$.
In this scheme, the form factors $F_{s,t,u,v}$ are defined generically as:
\begin{equation}
\label{eq:form}
F_{s}=\frac{\Lambda^{4}}{\Lambda^{4}+(s-M^{2}_{N})^{2}},
\,\,\,\,
F_{t}=\frac{\Lambda^{4}}{\Lambda^{4}+(t-M^{2}_{K})^{2}},
\,\,\,\,
F_{u}=\frac{\Lambda^{4}}{\Lambda^{4}+(u-M^{2}_{\Lambda^{*}})^{2}},
\,\,\,\,
F_{v}=\frac{\Lambda^{4}}{\Lambda^{4}+(t-M^{2}_{K^{*}})^{2}}.
\end{equation}
Here, $M_{s,t,u}$ are the masses of the off-shell particles in the $(s,t,u)$-channels. $\Lambda$ stands for a phenomenological cutoff parameter
determined by matching with experimental data. The {\it common} form factor $F_{c}$, which plays a crucial role to keep the gauge invariance, reads:
\begin{equation}
\label{eq:fc}
F_{c}=F_{s}+F_{t}-F_{s}F_{t}.
\end{equation}
It is clear that the $F_{c}$ satisfies the on-shell condition when one of $F_{s}$ and $F_{t}$ is on-shell. We note that there are several different
gauge-invariant form factors, as suggested in Ref.~\cite{Haberzettl:2006bn}. The choice of the scheme brings some uncertainty which
is numerically negligible.

\subsection{Resonance contribution}
There is little experimental information on the nucleon resonances coupling to $\Lambda^{*}$.
The situation is even worse for the hyperon resonances decaying
into $\gamma\Lambda^{*}$. Only some theoretical calculations have provided information on the
decays~\cite{Capstick:1998uh,Capstick:2000qj}. Unlike
the ground state $\Lambda(1116)$ photoproduction there nucleon and hyperon resonances play
important roles to reproduce the data~\cite{Janssen:2001pe}, the Born terms alone are enough to explain the available experimental data for the $\Lambda^{*}$ photorpdocution~\cite{Barber:1980zv,Muramatsu:2009zp}.
More dedicated experiments may show otherwise in the future.
Keeping this situation in mind, we attempt to include nucleon resonance contributions using the result of the relativistic constituent-quark model calculation~\cite{Capstick:1998uh}. Among the possible nucleon resonances given in Ref.~\cite{Capstick:1998uh}, we only choose $D_{13}(2080)$ with the two-star confirmation ($**$) but neglect $S_{11}(2080)$ and $D_{15}(2200)$ because that $S_{11}$ is still in poor confirmation ($*$), as for $D_{15}$, we do not have experimental data for the helicity amplitudes which are necessary to determine the strength of the transition $\gamma N\to N^{*}$~\cite{Amsler:2008zzb}. Moreover the spin $5/2$ Lorentz structure of $D_{15}$ bring theoretical uncertainties~\cite{Choi:2007gy}.
In the quark model of Ref.~\cite{Capstick:1998uh}, $N^{*}(1945,3/2^{-})$ is identified as $D_{13}$. However, we prefer to adopt the experimental value for the $D_{13}$ mass. Its transition and strong interactions are defined as follows:
\begin{eqnarray}
\label{eq:lreso}
\mathcal{L}_{\gamma NN^{*}}&=&
-\frac{ie_{Q}f_{1}}{2M_{N}}
\bar{N}^{*}_{\mu} \gamma_{\nu}F^{\mu\nu}N
-\frac{e_{Q}f_{2}}{(2M_{N})^{2}}
\bar{N}^{*}_{\mu}F^{\mu\nu}(\partial_{\nu}N)+\mathrm{h.c.},
\cr
\mathcal{L}_{KN^{*}\Lambda^{*}}&=&
\frac{g_{1}}{M_{K}}
\bar{\Lambda}^{*}_{\mu}\gamma_{5}\rlap{/}{\partial}
KN^{*\mu}+
\frac{ig_{2}}{M^{2}_{K}}\bar{\Lambda}^{*}_{\mu}\gamma_{5}
(\partial_{\mu}\partial_{\nu}
K)N^{*\nu}+\mathrm{h.c.},
\end{eqnarray}
where $N^{*}$ denotes the field for $D_{13}$. The coupling constants $f_{1}$ and $f_{2}$ can be computed using the helicity amplitudes~\cite{Oh:2007jd}:
\begin{eqnarray}
\label{eq:f1f2}
A^{p^{*}}_{1/2}&=&\frac{e_{Q}\sqrt{6}}{12}
\left(\frac{|\bm{k}_{\gamma N}|}{M_{D_{13}}M_{N}} \right)^{\frac{1}{2}}
\left[f_{1}+\frac{f_{2}}{4M^{2}_{N}}M_{D_{13}}(M_{D_{13}}+M_{N}) \right],
\cr
A^{p^{*}}_{3/2}&=&\frac{e_{Q}\sqrt{2}}{4M_{N}}
\left(\frac{|\bm{k}_{\gamma N}|M_{D_{13}}}{M_{N}} \right)^{\frac{1}{2}}
\left[f_{1}+\frac{f_{2}}{4M_{N}}(M_{D_{13}}+M_{N}) \right],
\end{eqnarray}
where the superscript $p^{*}$ indicates the positive-charge $D_{13}$, and $|\bm{k}_{\gamma N}|=828$ MeV in the decay process of $D_{13}\to\gamma N$ using Eq.~(\ref{eq:kollen}). The experimental values for $A^{p^{*}}_{1/2}$ and $A^{p^{*}}_{3/2}$ are taken from Ref.~\cite{Amsler:2008zzb}:
\begin{equation}
\label{eq:hel}
A^{p^{*}}_{1/2}=(-0.020\pm0.008)/\sqrt{\mathrm{GeV}},\,\,\,\,
A^{p^{*}}_{3/2}=(0.017\pm0.011)/\sqrt{\mathrm{GeV}}.
\end{equation}
We obtain $e_{Q}f_{1}=-0.19$ and $e_{Q}f_{2}=0.19$. The strong coupling strengths $g_{1}$ and $g_{2}$ are given by~\cite{Oh:2007jd}:
\begin{equation}
\label{eq:qqq}
G_{1}=G_{11}\frac{g_{1}}{M_{K}}+G_{12}\frac{g_{2}}{M^{2}_{K}},\,\,\,\,
G_{3}=G_{31}\frac{g_{1}}{M_{K}}+G_{32}\frac{g_{2}}{M^{2}_{K}}.
\end{equation}
Here the coefficients $G_{11,12,31,32}$ are defined as
\begin{eqnarray}
\label{eq:ggggg}
G_{11}&=&\frac{\sqrt{30}}{60\sqrt{\pi}}\frac{1}{M_{\Lambda^{*}}}
\left(\frac{|\bm{k}_{K\Lambda^{*}}|}{M_{D_{13}}} \right)^{\frac{1}{2}}
\sqrt{E_{\Lambda^{*}}-M_{\Lambda^{*}}}
(M_{D_{13}}+M_{\Lambda^{*}})(E_{\Lambda^{*}}+4M_{\Lambda^{*}}),
\cr
G_{12}&=&-\frac{\sqrt{30}}{60\sqrt{\pi}}\frac{|\bm{k}_{K\Lambda^{*}}|^{2}}{M_{\Lambda^{*}}}\sqrt{|\bm{k}_{\Lambda^{*}}|M_{D_{13}}}
\sqrt{E_{\Lambda^{*}}-M_{\Lambda^{*}}},
\cr
G_{31}&=&-\frac{\sqrt{30}}{20\sqrt{\pi}}\frac{1}{M_{\Lambda^{*}}}
\left(\frac{|\bm{k}_{K\Lambda^{*}}|}{M_{D_{13}}} \right)^{\frac{1}{2}}
\sqrt{E_{\Lambda^{*}}-M_{\Lambda^{*}}}
(M_{D_{13}}+M_{\Lambda^{*}})(E_{\Lambda^{*}}-M_{\Lambda^{*}}),
\cr
G_{32}&=&\frac{\sqrt{30}}{20\sqrt{\pi}}\frac{|\bm{k}_{K\Lambda^{*}}|^{2}}{M_{\Lambda^{*}}}\sqrt{|\bm{k}_{\Lambda^{*}}|M_{D_{13}}}
\sqrt{E_{\Lambda^{*}}-M_{\Lambda^{*}}},
\end{eqnarray}
where $|\bm{k}_{K\Lambda^{*}}|=224$ MeV in the decay process of $D_{13}\to K\Lambda^{*}$ and $E^{2}_{\Lambda^{*}}=M^{2}_{\Lambda^{*}}+\bm{k}^{2}_{K\Lambda^{*}}$. Employing the theoretical estimations on $G_{1}\approx-2.6\,\sqrt{\mathrm{MeV}}$ and $G_{3}\approx-0.2\,\sqrt{\mathrm{MeV}}$ given in Refs.~\cite{Capstick:1998uh,Capstick:2000qj}, one has $g_{1}=-1.07$ and $g_{2}=-3.75$.
Now the scattering amplitude for $D_{13}$ in the $s$-channel can be written as follows:
\begin{eqnarray}
\label{eq:re}
i\mathcal{M}^{*}_{s}&=&\bar{u}^{\mu}_{2}\gamma_{5}\Bigg\{
\frac{e_{Q}f_{1}g_{1}}
{2M_{K}M_{N}}\rlap{/}{k}_{2}
\left[\frac{(\rlap{/}{k}_{1}+\rlap{/}{p}_{1}+M_{D_{13}})
(k_{1\mu}\rlap{/}{\epsilon}-\rlap{/}{k}_{1}\epsilon_{\mu})}
{s-M^{2}_{D_{13}}-iM_{D_{13}}
\Gamma_{D_{13}}} \right]
\cr
&+&\frac{e_{Q}f_{2}g_{1}}
{4M_{K}M^{2}_{N}}\rlap{/}{k}_{2}
\left[\frac{(\rlap{/}{k}_{1}+\rlap{/}{p}_{1}+M_{D_{13}})
[k_{1\mu}(p_{1}\cdot\epsilon)-(p_{1}\cdot k_{1})\epsilon_{\mu}]}
{s-M^{2}_{D_{13}}-iM_{D_{13}}
\Gamma_{D_{13}}} \right]
\cr
&-&\frac{e_{Q}f_{1}g_{2}}
{2M^{2}_{K}M_{N}}k_{2\mu}
\left[\frac{(\rlap{/}{k}_{1}+\rlap{/}{p}_{1}+M_{D_{13}})
[(k_{1}\cdot k_{2})\rlap{/}{\epsilon}-\rlap{/}{k}_{1}(\epsilon\cdot k_{2})]}
{s-M^{2}_{D_{13}}-iM_{D_{13}}
\Gamma_{D_{13}}} \right]
\cr
&-&\frac{e_{Q}f_{2}g_{2}}
{4M^{2}_{K}M^{2}_{N}}k_{2\mu}
\left[\frac{(\rlap{/}{k}_{1}+\rlap{/}{p}_{1}+M_{D_{13}})
[(k_{1}\cdot k_{2})(\epsilon\cdot p_{2})-(\epsilon\cdot k_{2})(k_{1}\cdot p_{2})]}
{s-M^{2}_{D_{13}}-iM_{D_{13}}
\Gamma_{D_{13}}} \right]\Bigg\}u_{1}F_{s},
\end{eqnarray}
where $\Gamma_{D_{13}}$ is the full decay width and has large experimental uncertainty, $\Gamma_{D_{13}}=(87\sim1075)$ MeV~\cite{Amsler:2008zzb}. The preferred values for the PDG average locate at $(180\sim 450)$ MeV. Considering this situation, as a trial, we choose $\Gamma_{D_{13}}\approx 500$ MeV. Actually there are only small differences in the numerical results even for sizable changes in $\Gamma_{D_{13}}$. We find that the $D_{13}$ resonance contribution will become pronounced
only if $\Gamma_{D_{13}}$ becomes far narrower as lower than $\sim$ 100 MeV. However, such a narrow nucleon resonance is unlikely to exist unless there are unusual production mechanisms for the resonance such as the exotics. Therefore through this article we keep $\Gamma_{D_{13}}=500$ MeV.

\subsection{Regge contributions}
In this subsection, we explain how the Regge contributions are implemented in the $\Lambda^{*}$ photoproduction. As done in Refs.~\cite{Corthals:2005ce,Corthals:2006nz,Ozaki:2009wp}, considering the pseudoscalar and vector strange-meson Regge trajectories, we replace the $K$ and $K^{*}$ propagators in Eq.~(\ref{eq:AMP}) as follows:
\begin{eqnarray}
\label{eq:RT}
\frac{1}{t-M^{2}_{K}}\to\mathcal{D}_{K}
&=&\left(\frac{s}{s_{0}} \right)^{\alpha_{K}}
\frac{\pi\alpha'_{K}}{\Gamma(1+\alpha_{K})\sin(\pi\alpha_{K})},
\cr
\frac{1}{t-M^{2}_{K^{*}}}\to
\mathcal{D}_{K^{*}}&=&\left(\frac{s}{s_{0}} \right)^{\alpha_{K^{*}}-1}
\frac{\pi\alpha'_{K}}{\Gamma(\alpha_{K})\sin(\pi\alpha_{K})}.
\end{eqnarray}
Here $\alpha'_{K,K^{*}}$ indicate the slopes of the trajectories. $\alpha_{K}$ and $\alpha_{K^{*}}$ are the linear trajectories of the mesons for even and odd spins, respectively, given as functions of $t$ assigned as
\begin{equation}
\label{eq:TR}
\alpha_{K}=0.70\,\mathrm{GeV}^{-2}(t-M^{2}_{K}),
\,\,\,\,
\alpha_{K^{*}}=1+0.85\,\mathrm{GeV}^{-2}(t-M^{2}_{K^{*}}).
\end{equation}
Here is a caveat; in deriving Eq.~(\ref{eq:TR}), all the even and odd spin trajectories are assumed to be degenerate, although in reality the vector-kaon trajectories are not degenerated~\cite{Corthals:2005ce,Corthals:2006nz}. Moreover, for convenience, we have set the phase factor for the Reggeized propagators to be positive unity as done in Ref.~\cite{Ozaki:2009wp}. The cutoff parameter $s_{0}$ is chosen to be $1$ GeV~\cite{Corthals:2005ce,Corthals:2006nz}. Hereafter, we use a notation $i\mathcal{M}^{\mathrm{Regge}}$ for the amplitude with the Reggeized propagators in Eq.~(\ref{eq:RT}).

If we employ these Reggeized propagators in Eq.~(\ref{eq:RT}) for the invariant amplitude in Eq.~(\ref{eq:AMP}), the gauge invariance is broken. Fortunately, the $K^{*}$-exchange contribution is not affected since it is gauge invariant by itself according to the antisymmetric tensor structure: $k_{1}\cdot(i\mathcal{M}^{\mathrm{Regge}}_{K^{*}})=0$. Hence, it is enough to consider the $K$-exchange, electric $s$-channel, and contact-term contributions which are all proportional to $F_{c}$ as shown in Eq.~(\ref{eq:AMP}). This situation can be represented by
\begin{equation}
\label{eq:WT}
k_{1}\cdot(i\mathcal{M}^{\mathrm{Regge}}_{K}+i\mathcal{M}^{E}_{s}+i\mathcal{M}_{c})\ne0,
\end{equation}
resulting in the breakdown of gauge invariance of the scattering amplitude. To remedy this problem, we redefine the relevant amplitudes as follows~\cite{Corthals:2005ce,Corthals:2006nz}:
\begin{equation}
\label{eq:WT1}
i\mathcal{M}_{K}+i\mathcal{M}^{E}_{s}+i\mathcal{M}_{c}
\,\,\to\,\,
i\mathcal{M}^{\mathrm{Regge}}_{K}+(i\mathcal{M}^{E}_{s}
+i\mathcal{M}_{c})(t-M^{2}_{K})\mathcal{D}_{K}
=i\mathcal{M}^{\mathrm{Regge}}_{K}+i\bar{\mathcal{M}}^{E}_{s}
+i\bar{\mathcal{M}}_{c}.
\end{equation}
It is easy to show that Eq.~(\ref{eq:WT1}) satisfies the gauge invariance: $k_{1}\cdot(i\mathcal{M}^{\mathrm{Regge}}_{K}+i\bar{\mathcal{M}}^{E}_{s}+i\bar{\mathcal{M}}_{c})=0$.

\begin{figure}[ht]
\includegraphics[width=12cm]{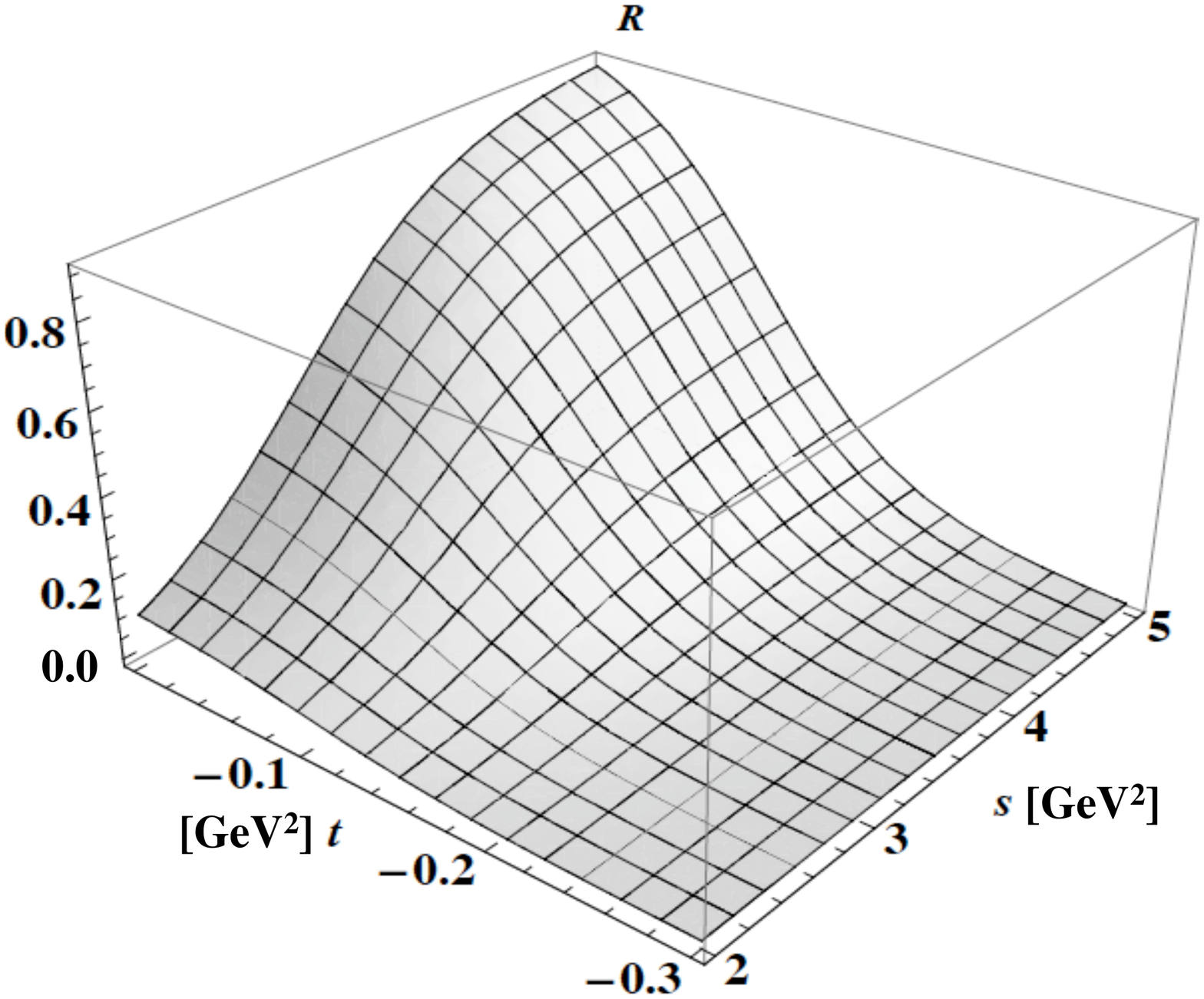}
\caption{$\mathcal{R}$ in Eq.~(\ref{eq:R}) as a function of $s$ and $|t|$ . We set $(s,t)_{\mathrm{Regge}}=(1,1)\,\mathrm{GeV}^{2}$ and $(s,t)_{0}=(1,1)\,\mathrm{GeV}^{2}$ as a trial.}
\label{FIG2}
\end{figure}

Considering that the Reggeized propagators work appropriately for $(s,|t|)\to(\infty, 0)$ and assume that the Regge contributions survive even in the low-energy region $(s,|t|)\to(s_{\mathrm{threshold}}, \mathrm{finite})$, it is natural to expect a smooth interpolation between two regions. The meson propagators are supposed to smoothly shifted from $\mathcal{D}_{K,K^{*}}$ for $s\gtrsim s_{\mathrm{Regge}}$ to a usual one for $s\lesssim s_{\mathrm{Regge}}$. Here, $s_{\mathrm{Regge}}$ indicates a certain value of $s$ from which the Regge contributions become effective. Similar consideration is also possible for $|t|$, and we can set $t_{\mathrm{Regge}}$ as well. Hence, as a trial, we parametrize the smooth interpolation by redefining the form factors in the relevant invariant amplitudes in Eq.~(\ref{eq:AMP}) as follows:
\begin{equation}
\label{eq:R}
F_{c,v}\to\bar{F}_{c,v}\equiv
\left[(t-M^{2}_{K,K^{*}})\mathcal{D}_{K,K^{*}}\right]
\mathcal{R}+F_{c,v}(1-\mathcal{R}),\,\,\,\,\mathcal{R}=\mathcal{R}_{s}\mathcal{R}_{t},
\end{equation}
where
\begin{equation}
\label{eq:RSRT}
\mathcal{R}_{s}=
\frac{1}{2}
\left[1+\tanh\left(\frac{s-s_{\mathrm{Regge}}}{s_{0}} \right) \right],\,\,\,\,
\mathcal{R}_{t}=
1-\frac{1}{2}
\left[1+\tanh\left(\frac{|t|-t_{\mathrm{Regge}}}{t_{0}} \right) \right].
\end{equation}
Here, $s_{0}$ and $t_{0}$ denote free parameters to make the arguments of $\tanh$ in Eq.~(\ref{eq:RSRT}) dimensionless. It is easy to understand that $\mathcal{R}_{s}$ goes to unity as $s\to\infty$ and zero as $s\to0$ around $s_{\mathrm{Regge}}$, and $\mathcal{R}_{t}$ zero as $|t|\to\infty$ and unity as $|t|\to0$ around $t_{\mathrm{Regge}}$. These asymptotic behaviors of $\mathcal{R}_{s}$ and $\mathcal{R}_{t}$ ensure that $\bar{F}_{c,v}$ in Eq.~(\ref{eq:R}) interpolate the two regions smoothly as shown in Figure~\ref{FIG2} where we plot $\mathcal{R}$ as a function of $s$ and $|t|$, showing that $\mathcal{R}$ approaches to unity as $s\to\infty$ and $|t|\to0$ with arbitrary choices for $(s,t)_{\mathrm{Regge}}=(s,t)_{0}=(1,1)\,\mathrm{GeV}^{2}$. We will determine the parameters, $(s,t)_{\mathrm{Regge}}$ and $(s,t)_{0}$, with experimental data in the next Section.

\section{Numerical Results}
Here we present our numerical results. First, we label two models in Table~\ref{table1}.
The model A represent our full calculation includes the Regge contributions.
The model B includes the Born diagrams and the nucleon-resonance contribution from $D_{13}$ only.
Throughout this article, the numerical results from the model A will be represented by solid lines, whereas dashed lines will be for the model B.

There are several free parameters in our model. One is the vector-kaon coupling constant $g_{K^{*}N\Lambda^{*}}$. Its value was determined from the unitarized chiral model~\cite{Hyodo:2006uw}.
It is considerably smaller than $g_{KN\Lambda^{*}}$. Actually the experimental data from Ref.~\cite{Muramatsu:2009zp} showed that the $K^{*}$-exchange contribution must be far smaller than that of the contact term. As discussed in Refs. ~\cite{Nam:2005uq,Nam:2006cx,Nam:2009cv}, the effect from the $K^{*}$-exchange contribution with various choices of $g_{K^{*}N\Lambda^{*}}$ turns out to be not so essential.

In contrast, we note that the $K^{*}-$ exchange contributes significantly to the photon-beam asymmetry ($\Sigma$) ~\cite{Nam:2006cx}.
The experimental data of $\Sigma$ are given in Refs.~\cite{Kohri:2009xe,Muramatsu:2009zp}. For $\theta\lesssim60^{\circ}$, the value of $\Sigma$ was estimated to be $-0.01\pm0.07$, indicating that $g_{K^{*}N\Lambda^{*}}$ is small~\cite{Muramatsu:2009zp} compared with that given in Ref.~\cite{Nam:2006cx}. In Ref.~\cite{Kohri:2009xe}, it was also measured that $-0.1\lesssim\Sigma\lesssim0.1$ for $1.75\,\mathrm{GeV}\le E_{\gamma}\le2.4\,\mathrm{GeV}$, and this result supports $g_{K^{*}N\Lambda^{*}}\ll1$.

Taking into account all of these experimental and theoretical results, it is safe to set $g_{K^{*}N\Lambda^{*}}\approx0$.
There is a similar hybrid approach \cite{Toki08} based on the quark-gluon string mechanism in high energy region. They also found that
the K* exchange in t-channel is very small compared with the K exchange.
Thus, we will drop the $K^{*}$-exchange contribution from now on. Similarly, as shown in Ref.~\cite{Nam:2005uq}, the different choices of the anomalous magnetic moment of $\Lambda^{*}$, $\kappa_{\Lambda^{*}}$, does not make any significant numerical impact since the $u$-channel contribution is suppressed by the form factor $F_{u}$ in Eq.~(\ref{eq:form}). Hence, we will set $\kappa_{\Lambda^{*}}$ to be zero hereafter.
\begin{table}[b]
\begin{tabular}{c||c|c|c|c|c|c|c}
&$s$ channel&$u$ channel&$t_{K}$ channel&$t_{K^{*}}$ channel& contact term&$D_{13}$ resonance&Regge\\
\hline
Model A&$i\bar{\mathcal{M}}^{E}_{s}$,$i\mathcal{M}^{M}_{s}$&$i\mathcal{M}^{M}_{u}$&$i\mathcal{M}^{\mathrm{Regge}}_{K}$&$iM^{\mathrm{Regge}}_{K^{*}}$&$i\bar{\mathcal{M}}_{c}$&$i\mathcal{M}^{*}_{s}$
&$\mathcal{R}=\mathcal{R}_{s}\mathcal{R}_{t}$\\
Model B&$i\mathcal{M}^{E}_{s}$,$i\mathcal{M}^{M}_{s}$&$i\mathcal{M}^{M}_{u}$&$i\mathcal{M}_{K}$&$i\mathcal{M}_{K^{*}}$&$i\mathcal{M}_{c}$&$iM^{*}_{s}$&$\mathcal{R}=0$
\end{tabular}
\caption{Relevant amplitudes in the model A and B.}
\label{table1}
\end{table}

\subsection{Angular dependence}
\begin{figure}[ht]
\begin{tabular}{cc}
\includegraphics[width=7.5cm]{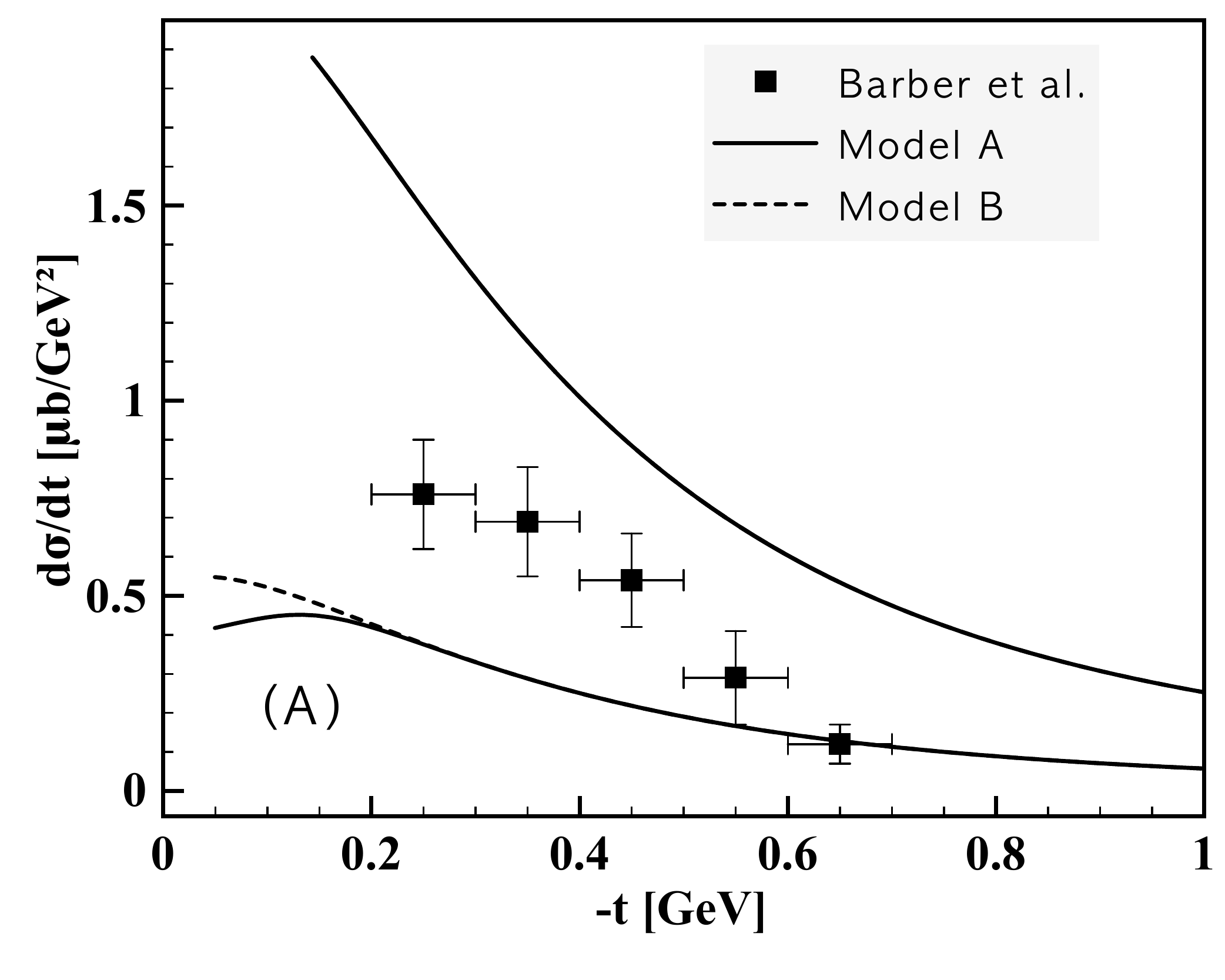}
\includegraphics[width=7.5cm]{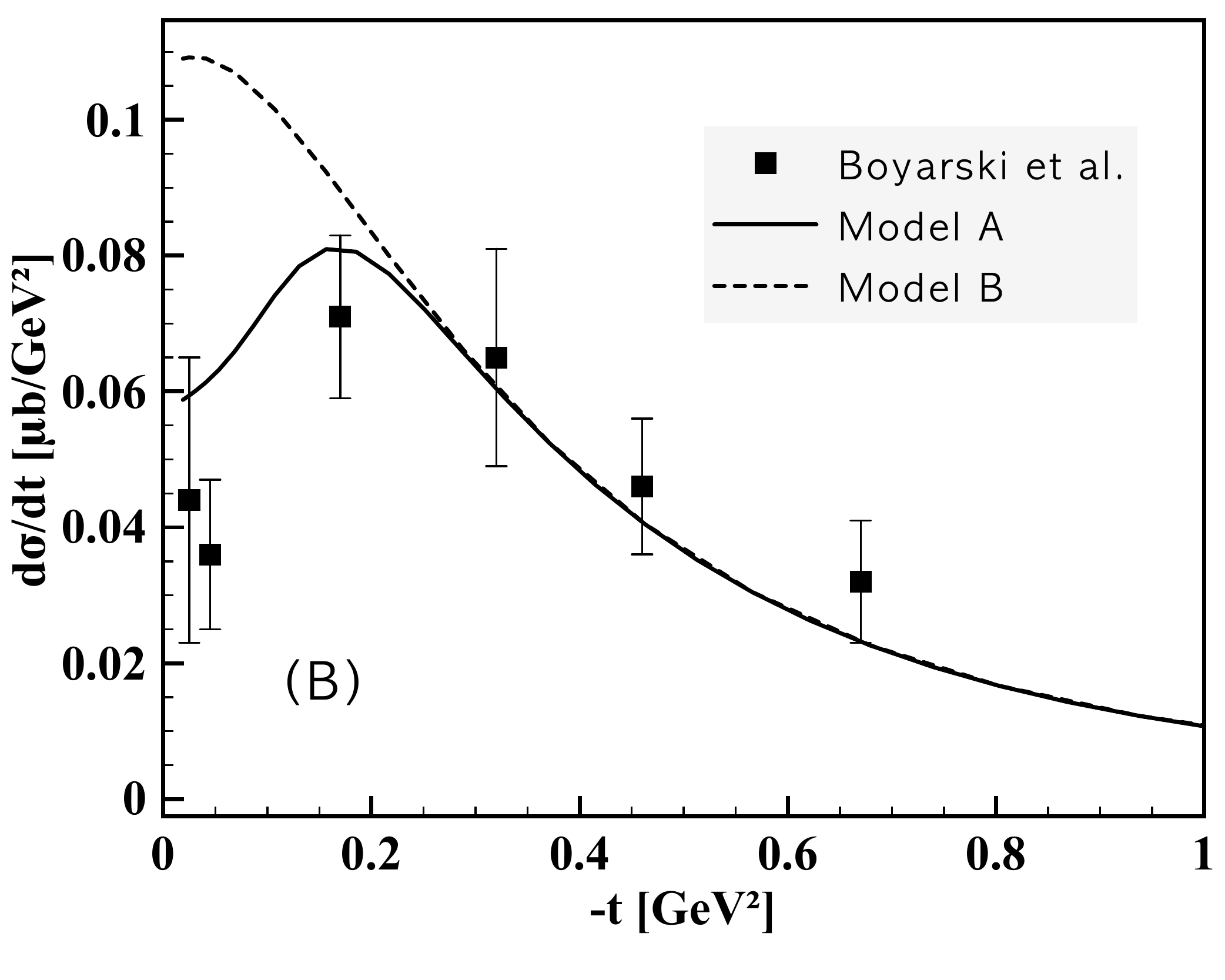}
\end{tabular}
\caption{Momentum transfer in the $t$ channel, $d\sigma/dt$ as a function of $-t$ for $E_{\gamma}=(2.4\sim4.8)$  GeV (A) and $E_{\gamma}=11$ GeV (B). The model A and B are explained in the text. The experimental data are taken from Ref.~\cite{Barber:1980zv,Boyarski:1970yc}.}
\label{FIG3}
\end{figure}

We first study $d\sigma/dt$ for the low- and high-energy experiments~\cite{Barber:1980zv,Boyarski:1970yc}. The  experiments were performed at $E_{\gamma}=(2.4\sim4.8)$ GeV~\cite{Barber:1980zv} and $E_{\gamma}=11$ GeV~\cite{Boyarski:1970yc}. In (A) of Figure~\ref{FIG3} the numerical results for $d\sigma/dt$ for the model A and B are almost identical except for very small $|t|$ regions, and reproduce the data qualitatively well. This observation indicates that the Regge contributions is
very small in the low-energy region as expected. In contrast, as in (B) of Figure~\ref{FIG3} with the data taken from Ref.~\cite{Boyarski:1970yc} (high energy), the results of the model A and B are very different. Note that a sudden change in the data around $|t|\approx0.2\,\mathrm{GeV}^{2}$ are reproduced by the model A but not by the model B. It shows that the smooth interpolation of the Regge contributions given by Eq.~(\ref{eq:R}) is necessary to explain the experimental data.

The parameters for $(s,t)_{\mathrm{Regge}}$ and $(s,t)_{0}$ employed for drawing the curves in Figure~\ref{FIG3}, are listed in Table~\ref{table2}. Here, we have chosen $s_{\mathrm{Regge}}=9\,\mathrm{GeV}^{2}$ which means that the Regge contributions become important for $\sqrt{s}>3$ GeV. The value of $t_{\mathrm{Regge}}$ is rather arbitrary. We set $t_{\mathrm{Regge}}=0.1\,\mathrm{GeV}^{2}$ because the physical situation changes drastically around this value. The other parameters, $(s,t)_{0}$ have been fixed to reproduce the data. We will adopt these values hereafter. Although the data of Ref.~\cite{Boyarski:1970yc} in the vicinity of $|t|\approx0$ show a decrease with respect to $|t|$, we did not fine tune our parameters for it because of the large experimental errors and qualitative nature of this work. We also find that the $D_{13}$ contribution is almost negligible as long as we use the input discussed in the previous section.

\begin{figure}[ht]
\includegraphics[width=16cm]{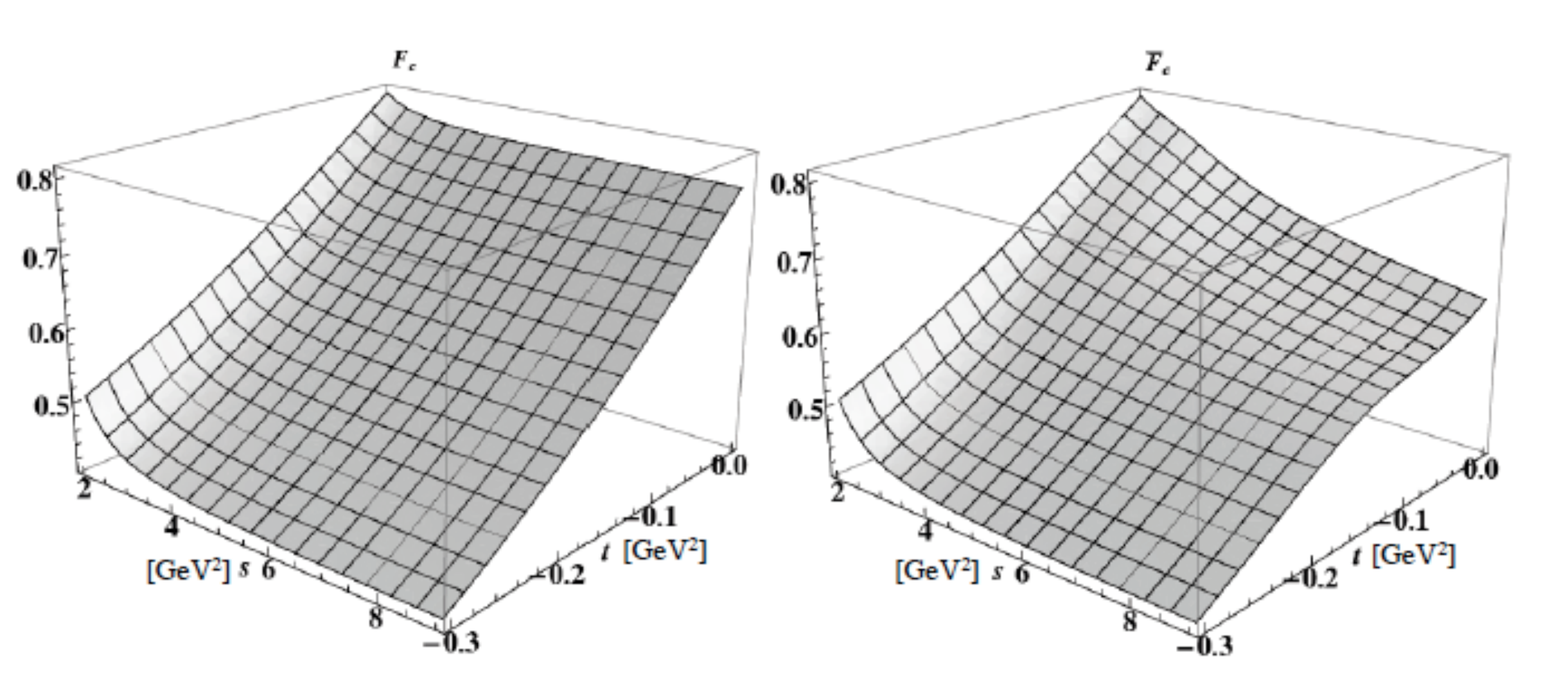}
\caption{$F_{c}$ (left) and $\bar{F}_{c}$ (right) as functions of $s$ and $t$, using the parameters listed in Table~\ref{table2}.}
\label{FIG4}
\end{figure}

Now, we want to take a close look on the bump structure around $|t|\approx0.2\,\mathrm{GeV}^{2}$ shown in (B) of Figure~\ref{FIG3}. Since the angular dependences of the cross section is largely affected by the common form factor, it is instructive to show $F_{c}$ (left) and $\bar{F}_{c}$ (right) as functions of $s$ and $t$ in Figure~\ref{FIG4} with the parameters listed in Table~\ref{table2}. In the vicinity of small $|t|\lesssim0.2\,\mathrm{GeV}^{2}$ and large $s\gtrsim4\,\mathrm{GeV}^{2}$, the difference between two form factors become obvious, i.e. $F_{c}$ increases with respect to $|t|$ monotonically, while $\bar{F}_{c}$ shows a complicated structure as we approach small $|t|$ region. Moreover, we can clearly see a bump-like structure around  $|t|\approx(0.1\sim0.2)\,\mathrm{GeV}^{2}$ at large $s$ region. This behavior of $\bar{F}_{c}$ cause the bump observed in the results for $d\sigma/dt$ as depicted in Figure~\ref{FIG3}. In other words, this structure is due to the Regge contributions for the  high $E_{\gamma}$ region indeed. Hence we conclude that the present reaction process is still dominated by the contact-term contribution as in ~\cite{Nam:2005uq,Nam:2006cx}, and the Regge contributions modify it in the vicinity near $|t|\lesssim0.2\,\mathrm{GeV}^{2}$ beyond the resonance region.

\begin{table}[b]
\begin{tabular}{c|c|c|c|c}
$\Lambda_{\mathrm{A,B}}$
&$s_{\mathrm{Regge}}$&$s_{0}$&$t_{\mathrm{Regge}}$&$t_{0}$\\
\hline
$675$ GeV&$3.0\,\mathrm{GeV}^{2}$&
$1.0\,\mathrm{GeV}^{2}$&
$0.1\,\mathrm{GeV}^{2}$&
$0.08\,\mathrm{GeV}^{2}$
\end{tabular}
\caption{Cutoff mass for the model A and B, and input parameters for the function $\mathcal{R}$ in Eqs.~(\ref{eq:R}) and (\ref{eq:RSRT}).}
\label{table2}
\end{table}

\begin{figure}[ht]
\begin{tabular}{cc}
\includegraphics[width=7.5cm]{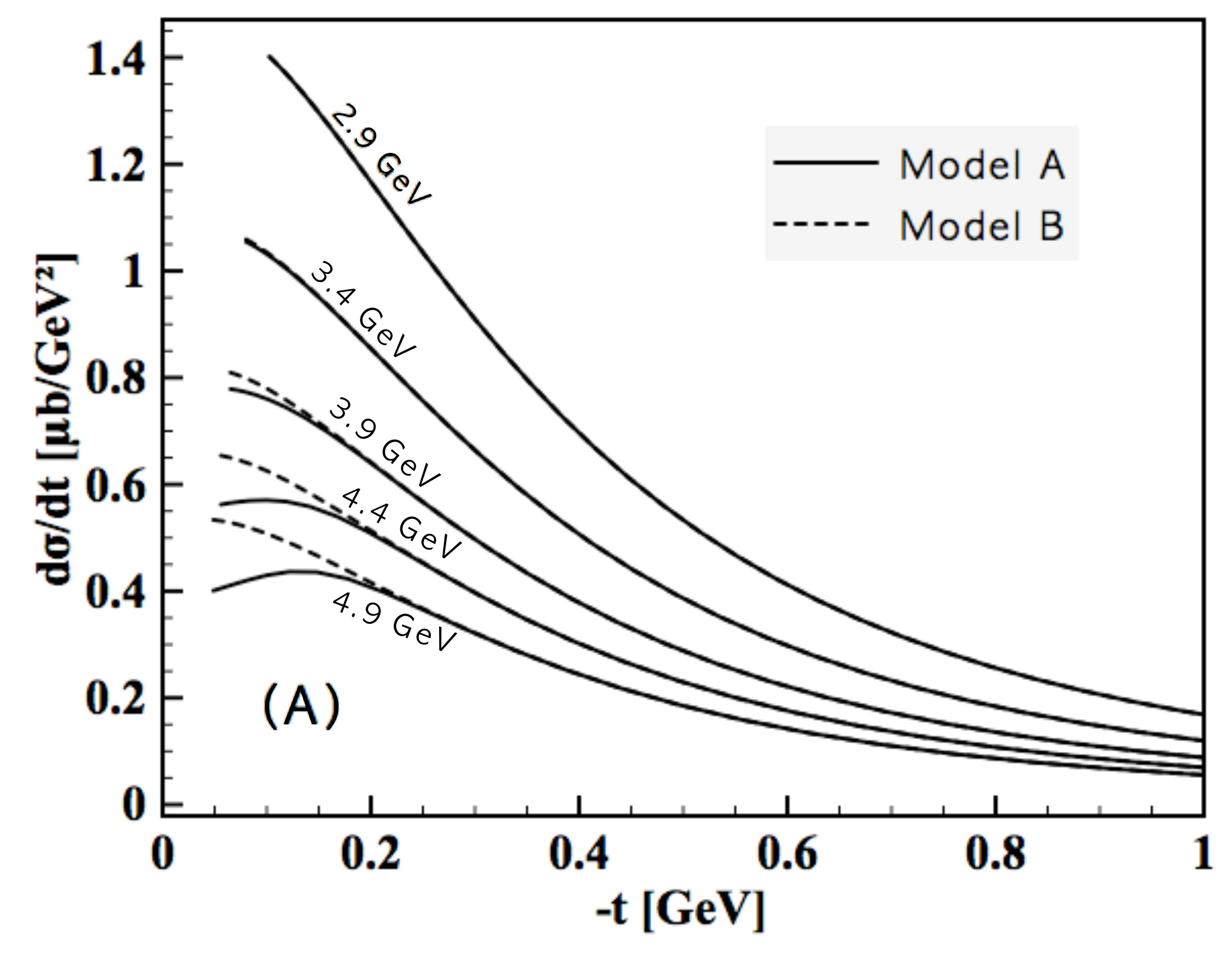}
\includegraphics[width=7.5cm]{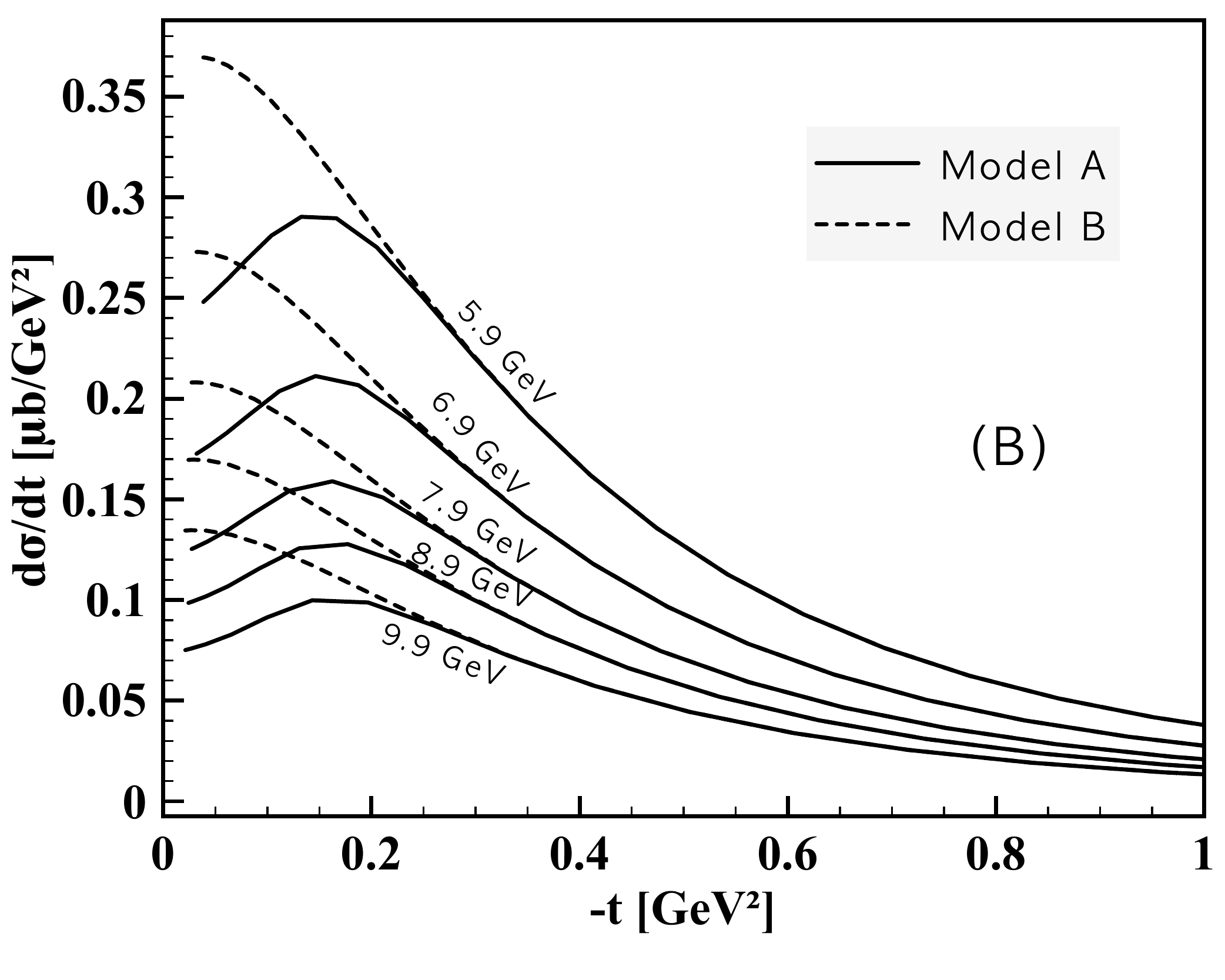}
\end{tabular}
\caption{Momentum transfer in the $t$ channel, $d\sigma/dt$ as a function of $-t$ for $E_{\gamma}=2.9$, $3.4$, $3.9$, $4.4$, and $4.9$ GeV (A). In (B), we draw it for $E_{\gamma}=5.9$, $6.9$, $7.9$, $8.9$, and $9.9$ GeV. The model A and B are explained in the text.}
\label{FIG5}
\end{figure}

\begin{figure}[ht]
\begin{tabular}{cc}
\includegraphics[width=7.5cm]{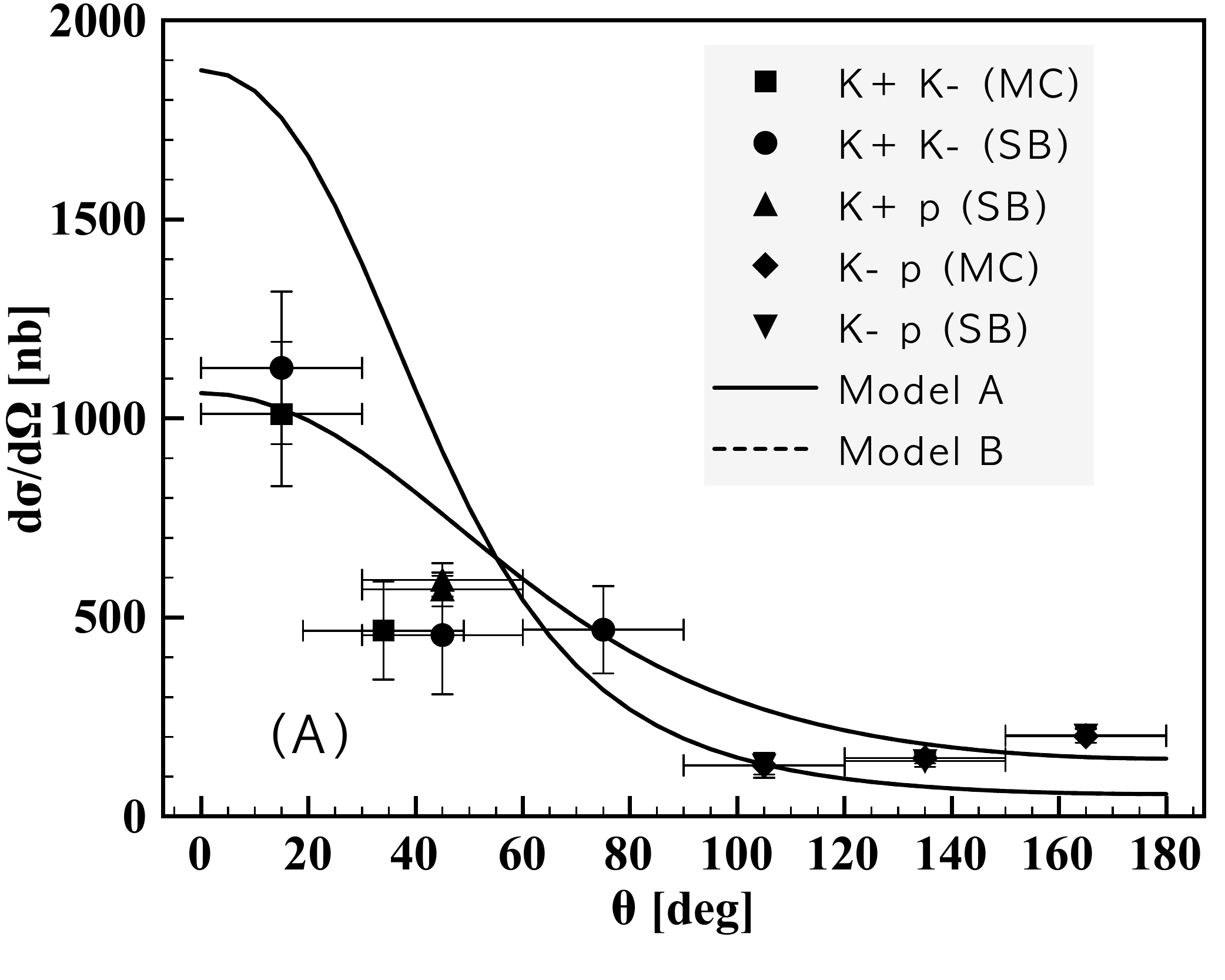}
\includegraphics[width=7.5cm]{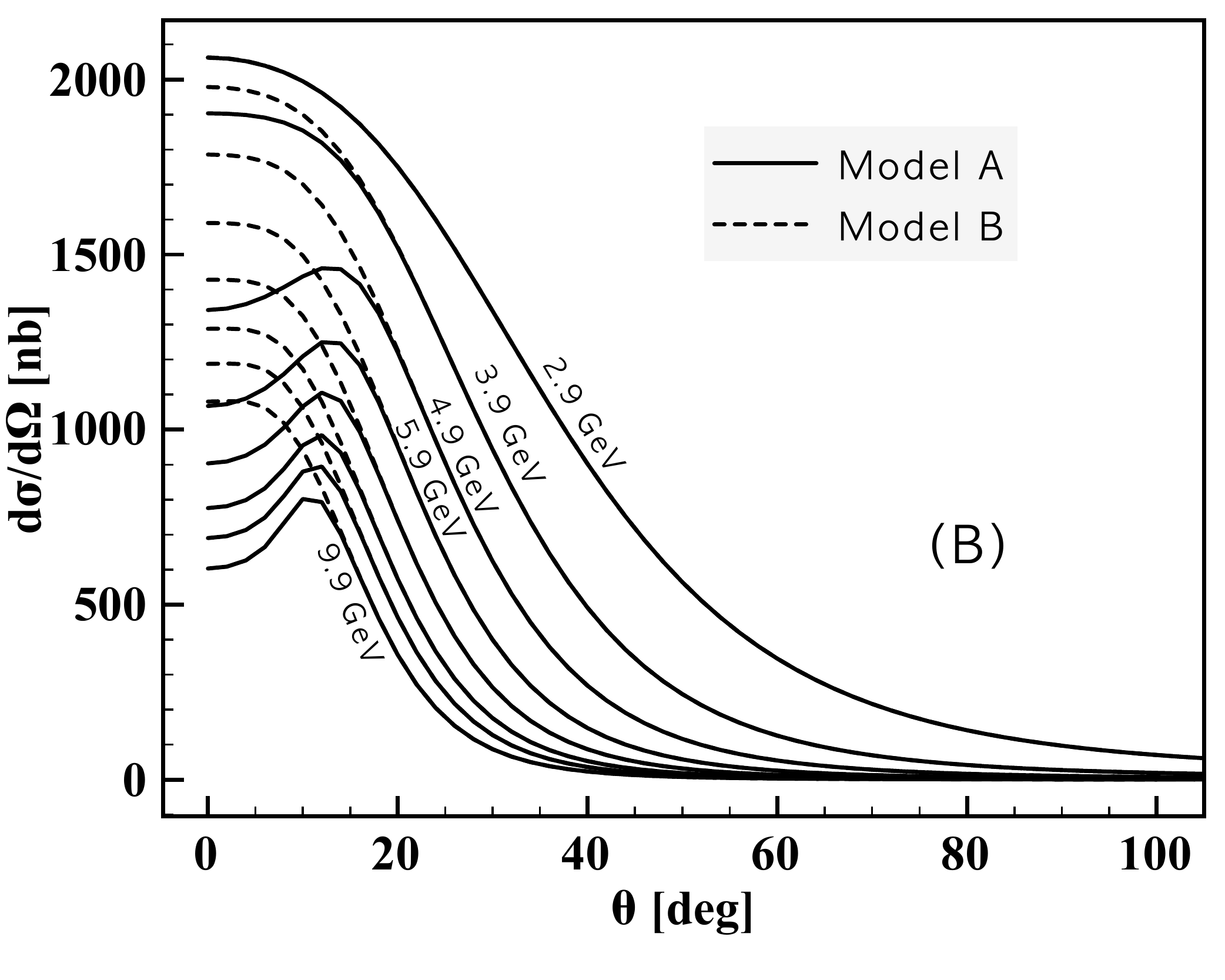}
\end{tabular}
\caption{Differential cross section, $d\sigma/d\Omega$ as a function of $\theta$ for $E_{\gamma}=(1.9\sim2.4)$ GeV, shown in (A). The experimental data are taken from Ref.~\cite{Muramatsu:2009zp}. In (B), we plot it for $E_{\gamma}=(2.9\sim9.9)$ GeV. The model A and B are explained in the text.}
\label{FIG6}
\end{figure}

In Figure~\ref{FIG5}, we depict the numerical results for
$d\sigma/dt$ as a function of $-t$ for the low- (A) and high-energy
(B) regions for more various energies. One finds that the Regge
contributions become visible beyond $E_{\gamma}\approx4$ GeV in the
small $|t|$ region. As the photon energy increases the bumps emerge
at $|t|=(0.1\sim0.2)\,\mathrm{GeV}^{2}$, indicating the effects from
the Regge contributions. In Figure~\ref{FIG6}, we plot
$d\sigma/d\Omega$ as a function of $\theta$. As given in (A), we
reproduce the experimental data qualitatively well for
$E_{\gamma}=(1.9\sim2.4)$ GeV, which represents the range of the
photon energy of LEPS collaboration~\cite{Muramatsu:2009zp}, showing
only negligible effects from the Regge contributions. The notations
for the data correspond to the channels ($K\bar{K}$, $KN$, and
$\bar{K}N$) in the $\gamma N\to K\bar{K}N$ reaction process and
analyzing methods (SB: side band and and MC: Monte Carlo). (B) of
the figure 6 shows the high-energy behavior of $d\sigma/d\Omega$ for
$E_{\gamma}=(2.9\sim9.9)$ GeV. The Regge contributions become
obvious beyond $E_{\gamma}\approx4$ GeV, and the bump in the
vicinity $\theta\approx10^{\circ}$ becomes narrower as $E_{\gamma}$
increases. The result od the model (B) is not shown here because it
is almost identical to the one of model (A).

\subsection{Energy dependence}
\begin{figure}[ht]
\includegraphics[width=7.5cm]{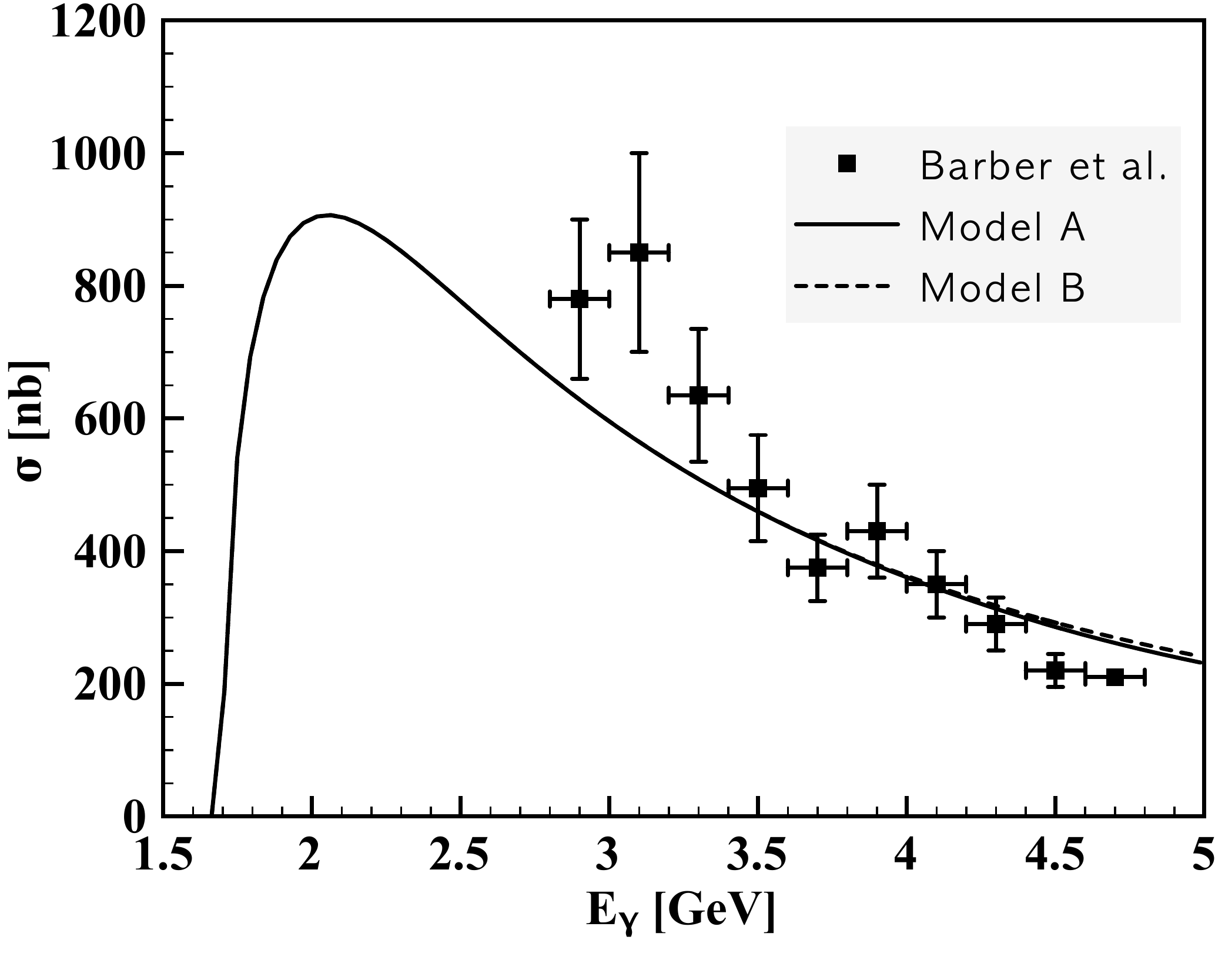}
\caption{Total cross section for the proton target as a function of $E_{\gamma}$. The model A and B are explained in the text.}
\label{FIG7}
\end{figure}

In Figure~\ref{FIG7} we present the numerical results for the total cross section as a function of $E_{\gamma}$ from the threshold to $E_{\gamma}=5$ GeV. We observe only small deviation between the model A and B beyond $E_{\gamma}\approx4$ GeV. It is consistent with the angular dependences as shown in the previous subsection. Obviously, there appear some unknown contributions at $E_{\gamma}\approx3$ GeV and $4$ GeV in the experimental data, which may correspond to nucleon or hyperon resonances not measured experimentally yet. For instance, at $E_{\gamma}\approx3$ corresponding to $\sqrt{s}\approx(2.5\sim2.6)$ GeV, $N^{*}(2600,11/2^{-})$ has been reported in Ref.~\cite{Amsler:2008zzb} with the $(***)$ confirmation. However, theoretical estimation for its coupling strength to $\Lambda^{*}$ is very small~\cite{Capstick:1998uh,Capstick:2000qj}.

\begin{figure}[ht]
\begin{tabular}{cc}
\includegraphics[width=7.5cm]{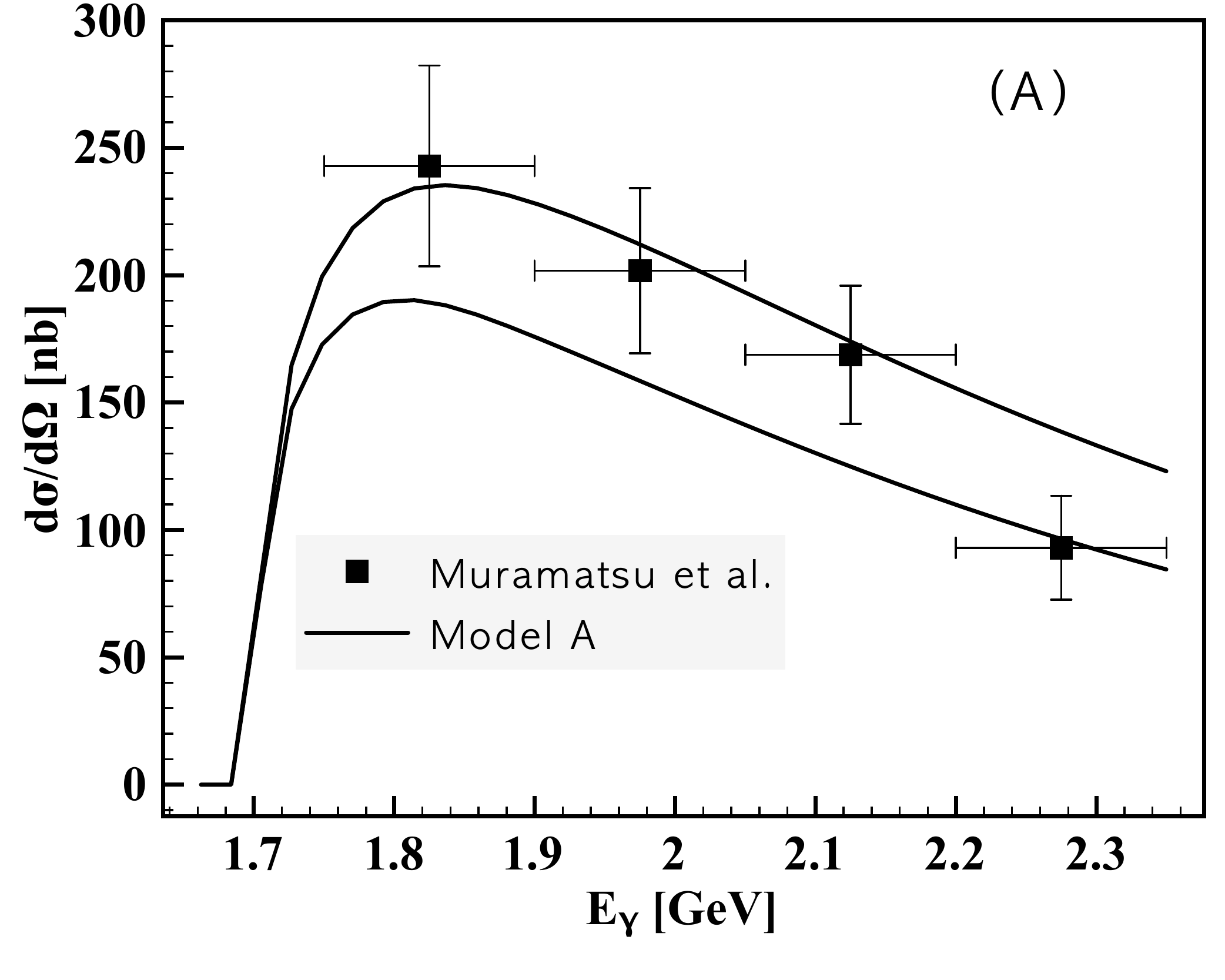}
\includegraphics[width=7.5cm]{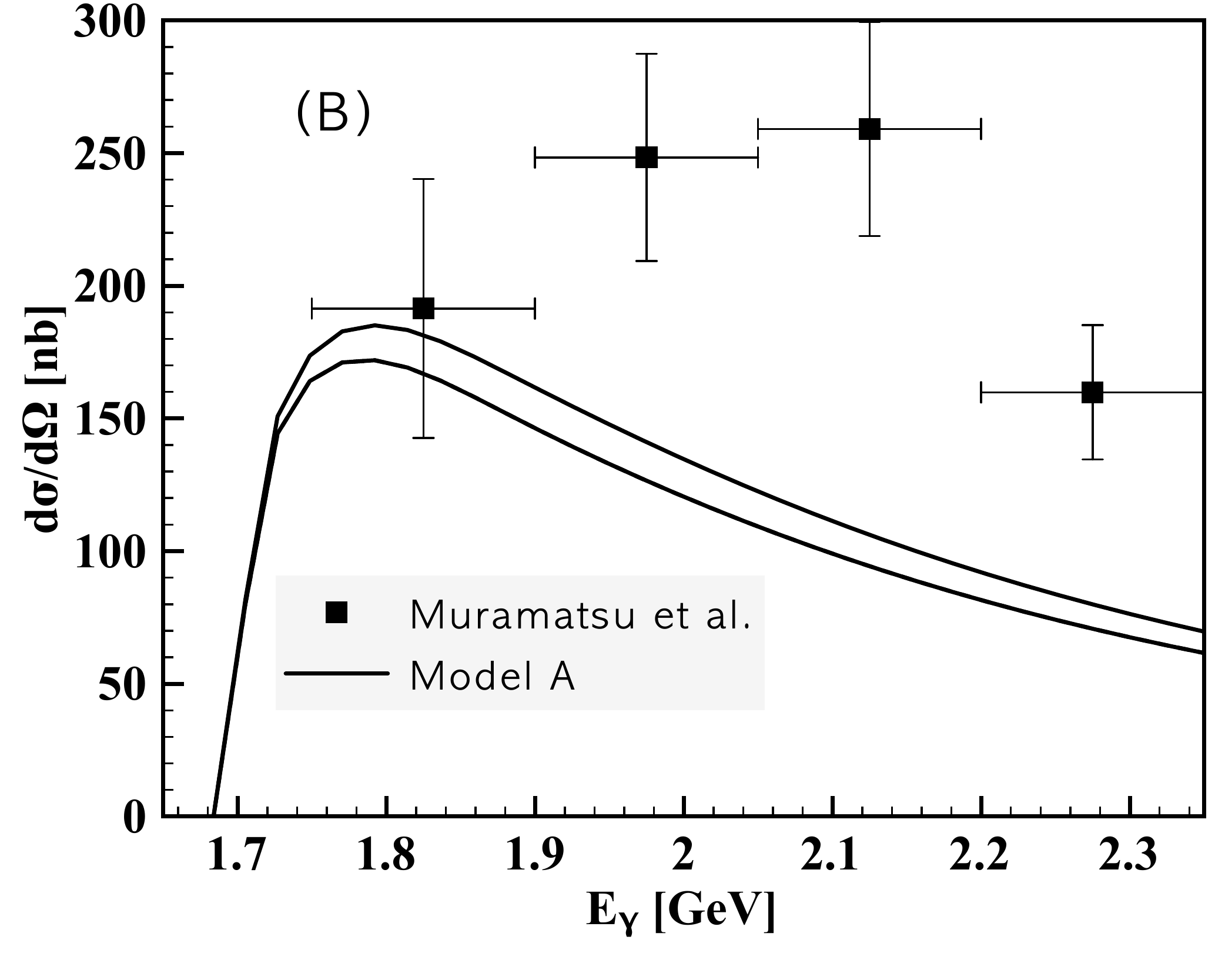}
\end{tabular}
\caption{Differential cross section, $d\sigma/d\Omega$ as a function of $E_{\gamma}$ for $\theta=(120\sim150)^{\circ}$ (A) and for $\theta=(150\sim180)^{\circ}$ (B). The experimental data are taken from Ref.~\cite{Muramatsu:2009zp}. The model A and B are explained in the text.}
\label{FIG8}
\end{figure}

In Figure~\ref{FIG8}, we plot $d\sigma/d\Omega$ as a function of $E_{\gamma}$ for $120^{\circ}\le\theta\le150^{\circ}$ (A) and for $\theta=(150\sim180)^{\circ}$ (B). The results of the model A and B coincide with each other because the Regge contribution is negligible in the low- energy region. Hence we only plot the numerical results of the model A. As for $\theta=(120\sim150)^{\circ}$ (A), the theoretical result reproduces the data qualitatively well, whereas it deviates much from the experimental data for $\theta=(150\sim180)^{\circ}$ (B). This deviation may signal a strong backward enhancement caused by unknown $u$-channel contributions which are not included in the present work. This strong backward enhancement is consistent with the increase in $d\sigma/d\Omega$ for $\theta=(100\sim180)^{\circ}$ as shown in (A) of Figure~\ref{FIG8}~\cite{Muramatsu:2009zp}. Although we do not show explicit results, we verified that, if we employ the simple Breit-Wigner form for a $u$-channel hyperon-resonance contribution as a trial, the increase in the backward region shown in (B) of Figure~\ref{FIG8} can be reproduced. However, it is a rather difficult to reproduce the data of $d\sigma/d\Omega$ in the backward direction simultaneously. Since we lack information on the interaction structure of the trial $u$-channel contribution, we will leave this task as a future work.

\subsection{Beam asymmetry}
\begin{figure}[ht]
\begin{tabular}{cc}
\includegraphics[width=7.5cm]{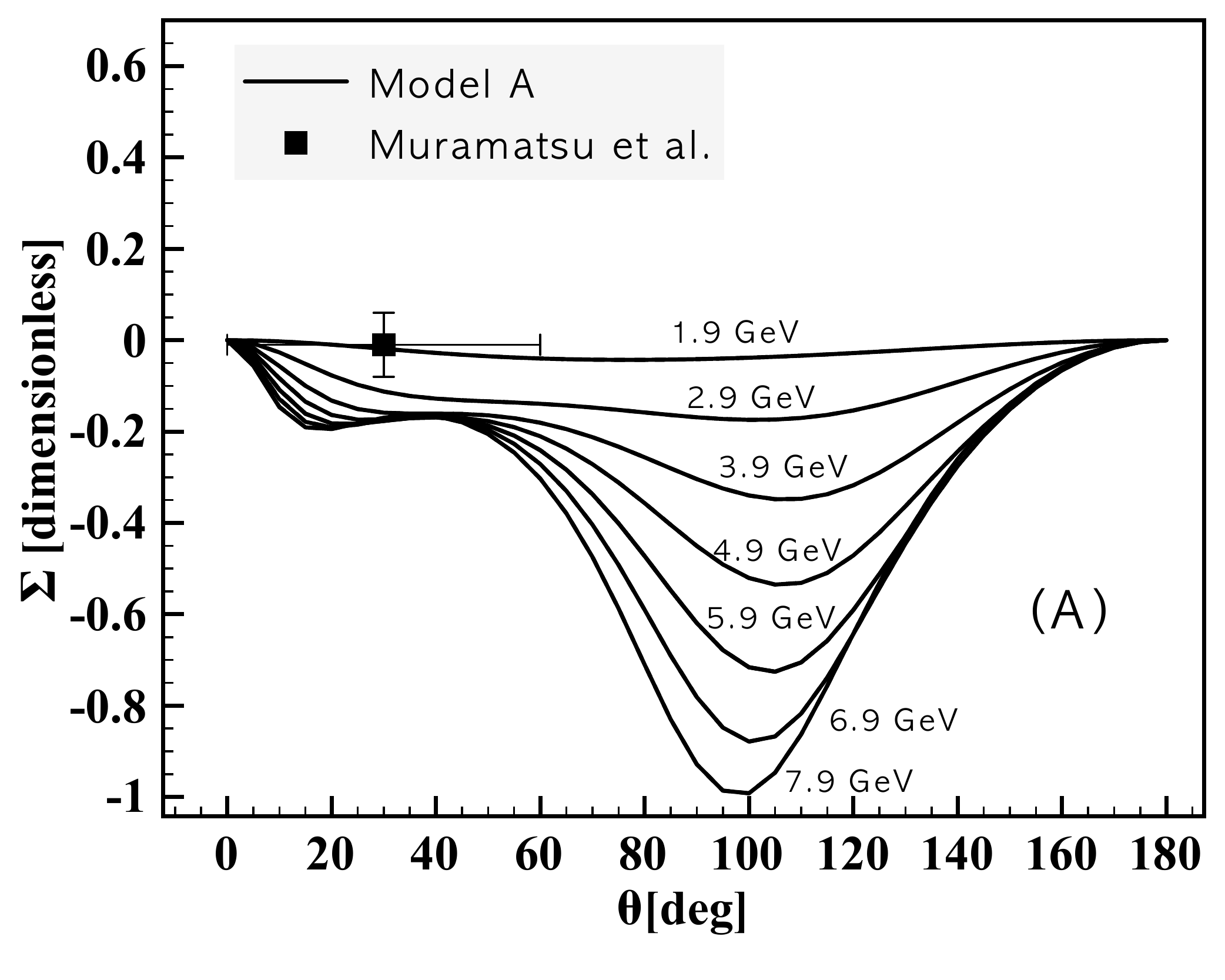}
\includegraphics[width=7.5cm]{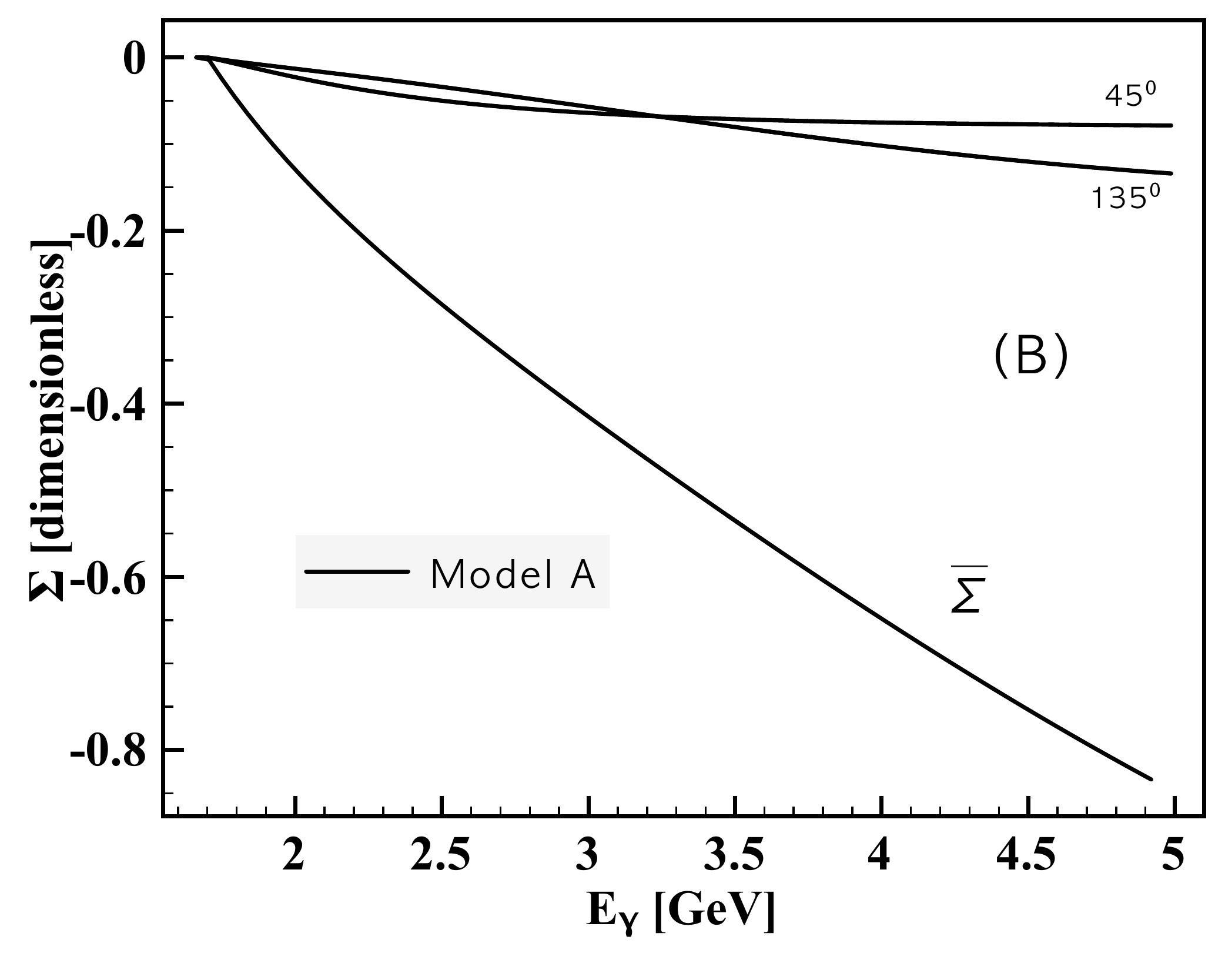}
\end{tabular}
\caption{Photon-beam asymmetry $\Sigma$ as a function of $\theta$ for $E_{\gamma}=(1.9\sim7.9)$ GeV (A). The data are taken from Ref.~\cite{Muramatsu:2009zp} for $E_{\gamma}=(1.75\sim2.4)$ GeV. In (B), we show $\Sigma$ as a function of $E_{\gamma}$ for $\theta=45^{\circ}$ and $135^{\circ}$, and $\bar{\Sigma}$ as in Eq.~(\ref{eq:IBA}).}
\label{FIG9}
\end{figure}

The photon-beam asymmetry defined in Eq.~(\ref{eq:BA}) can be measured in experiments using a linearly polarized photon beam:
\begin{equation}
\label{eq:BA}
\Sigma=\frac{\frac{d\sigma}{d\Omega}_{\perp}
-\frac{d\sigma}{d\Omega}_{\parallel}}
{\frac{d\sigma}{d\Omega}_{\perp}
+\frac{d\sigma}{d\Omega}_{\parallel}},
\end{equation}
where the subscripts $\perp$ and $\parallel$ denote the directions of the polarization which are perpendicular and parallel to the reaction plane, respectively. Here the reaction plane is defined by the $y$-$z$ plane, on which the incident photon along the $z$ direction and outgoing kaon reside. In (A) of Figure~\ref{FIG9}, we show $\Sigma$ as a function of $\theta$ for $E_{\gamma}=(1.9\sim7.9)$ GeV. The low-energy behavior is consistent with the previous work~\cite{Nam:2006cx}. As the energy increases, there appear a deeper valley around $\theta\approx100^{\circ}$. This behavior is mainly due to the $K$-exchange contribution, since it is enhanced with respect to $E_{\gamma}$ and contains a term $\propto k_{2}\cdot\epsilon$. It becomes zero for $d\sigma/d\Omega_{\perp}$ and finite for $d\sigma/d\Omega_{\parallel}$ resulting in $\Sigma\to-1$ as understood by Eq.~(\ref{eq:BA}). Hence, unlike the angular and energy dependences the photon-beam asymmetry is largely affected by the $K$-exchange contribution. Interestingly, the model A and B produce almost the same results for all the energies. Therefore, we will plot only the numerical results of the model A hereafter. Considering that the $s$- and $u$-channel are strongly suppressed in the present framework~\cite{Nam:2005uq,Nam:2006cx,Nam:2009cv} and the gauge invariance of the invariant amplitude, the invariant amplitude can be simplified as,
\begin{equation}
\label{eq:SIMAMP}
i\mathcal{M}_{\mathrm{total}}\approx(i\mathcal{M}_{c}+i\mathcal{M}^{E}_{s}+i\mathcal{M}_{t})\bar{F}_{c}.
\end{equation}
Hence, in general, the form factor $\bar{F}_{c}$ is factorized from the amplitude. In some quantity such as the ratio of the amplitude squared $\sim|\mathcal{M}_{\mathrm{total}}|^{2}$ its effect will be canceled. This cancelation occurs in the photon-beam asymmetry in Eq.~(\ref{eq:BA}).
In the Figure 9, we show the experimental data from Ref.~\cite{Muramatsu:2009zp}, in which $\Sigma$ was estimated $-0.01\pm0.07$ for $\theta=(0\sim60)^{\circ}$ for the LEPS photon-energy range $E_{\gamma}=(1.75\sim2.4)$ GeV. We note that the numerical results are in good agreement with the data. There is a strong experimental support for the assumption of $g_{K^{*}N\Lambda^{*}}\ll1$ for the proton-target case as mentioned already.

In (B) of Figure~\ref{FIG9}, we draw $\Sigma$ as a function of $E_{\gamma}$  for $\theta=45^{\circ}$ and $135^{\circ}$, and $\bar{\Sigma}$ defined as~\cite{Nam:2006cx},
\begin{equation}
\label{eq:IBA}
\bar{\Sigma}(E_{\gamma})=\frac{1}{2}\int^{\pi}_{0} \Sigma(\theta,E_{\gamma})\,\sin\theta\,d\theta,
\end{equation}
where the factor $1/2$ for normalization. As $E_{\gamma}$ increases, the absolute values of $\Sigma$ and $\bar{\Sigma}$ become larger, since the $K$-exchange contribution is enhanced with respect to $E_{\gamma}$ as discussed above.

\subsection{Polarization-transfer coefficients}
In this subsection, the polarization-transfer coefficients $C_{x}$ and $C_{z}$ for the $\Lambda^{*}$ photoproduction are presented. The $C_{x}$ and $C_{z}$ are identified as the spin asymmetry along the direction of the polarization of the recoil baryon with the circularly polarized photon beam. Physically, these quantities indicate how much the initial helicity transferred to the recoil baryon polarized in a certain direction. First, we define the polarization-transfer coefficients in the $(x',y',z')$ coordinate, being similar to those for the spin-$1/2$ hyperon photoproduction as in Refs.~\cite{McNabb:2003nf,Anisovich:2007bq}:
\begin{equation}
\label{eq:CXZ}
C_{x',|S_{x'}|}=
\frac{\frac{d\sigma}{d\Omega}_{r,0,+S_{x'}}
-\frac{d\sigma}{d\Omega}_{r,0,-S_{x'}}}
{\frac{d\sigma}{d\Omega}_{r,0,+S_{x'}}
+\frac{d\sigma}{d\Omega}_{r,0,-S_{x'}}},\,\,\,\,
C_{z',|S_{z'}|}=
\frac{\frac{d\sigma}{d\Omega}_{r,0,+S_{z'}}
-\frac{d\sigma}{d\Omega}_{r,0,-S_{z'}}}
{\frac{d\sigma}{d\Omega}_{r,0,+S_{z'}}
+\frac{d\sigma}{d\Omega}_{r,0,-S_{z'}}},
\end{equation}
where the subscripts $r$, $0$, and $\pm S_{x,'z'}$ stand for the right-handed photon polarization, unpolarized target nucleon, and polarization of the recoil baryon along the $x'$- or $z'$-axis, respectively. Since the photon helicity is fixed to be $+1$ here, the $C_{x'}$ and $C_{z'}$ measures the polarization transfer to the recoil baryon. Moreover, the $C_{x'}$ and $C_{z'}$ behave as the components of a three vector so that it can be rotated to the $(x,y,z)$ coordinate as:
\begin{equation}
\label{eq:ro}
\left(\begin{array}{c}
C_{x}\\C_{z}\end{array}\right)
=\left(
\begin{array}{cc}
\cos{\theta_{K}}&\sin{\theta_{K}}\\
-\sin{\theta_{K}}&\cos{\theta_{K}}
\end{array}
 \right)\left(\begin{array}{c}
C_{x'}\\C_{z'}\end{array}\right),
\end{equation}
where the $(x,y,z)$ coordinate stands for that the incident photon momentum is aligned to the $z$ direction. Being different from usual spin-$1/2$ baryon photoproductions, we will have four different polarization-transfer coefficients, $C_{x,1/2}$, $C_{z,1/2}$, $C_{x,3/2}$, and $C_{z,3/2}$, due to the total spin states of $\Lambda^{*}$. Note that, in terms of the helicity conservation, $C_{(x,z),1/2}$ and $C_{(x,z),3/2}$ should be zero and unity in the collinear limit ($\theta=0$ or $180^{\circ}$). More detailed discussions are given in Refs.~\cite{Nam:2009cv,Anisovich:2007bq,Schumacher:2008xw,Fasano:1992es,Artru:2006xf}.

\begin{figure}[ht]
\begin{tabular}{cc}
\includegraphics[width=7.5cm]{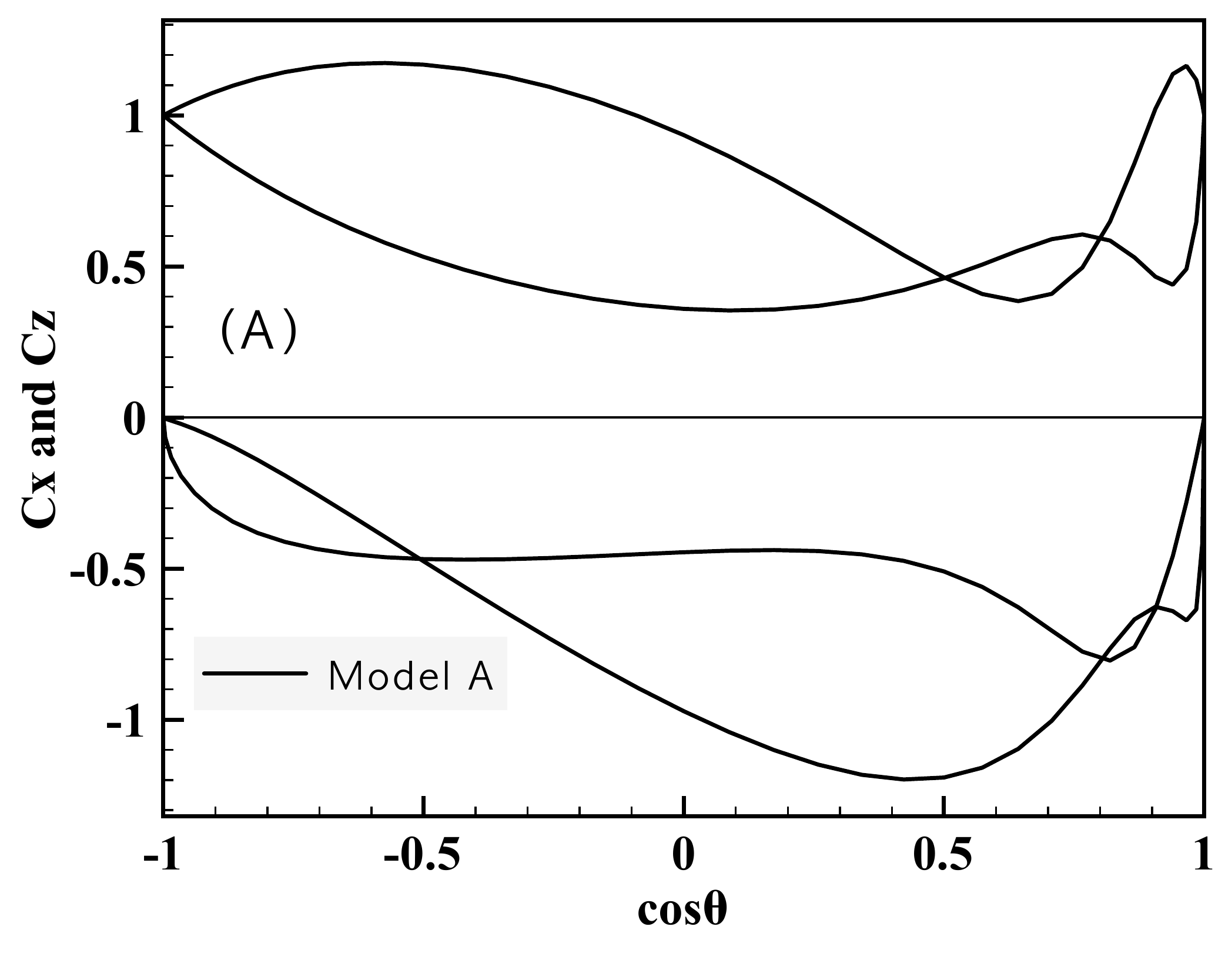}
\includegraphics[width=7.5cm]{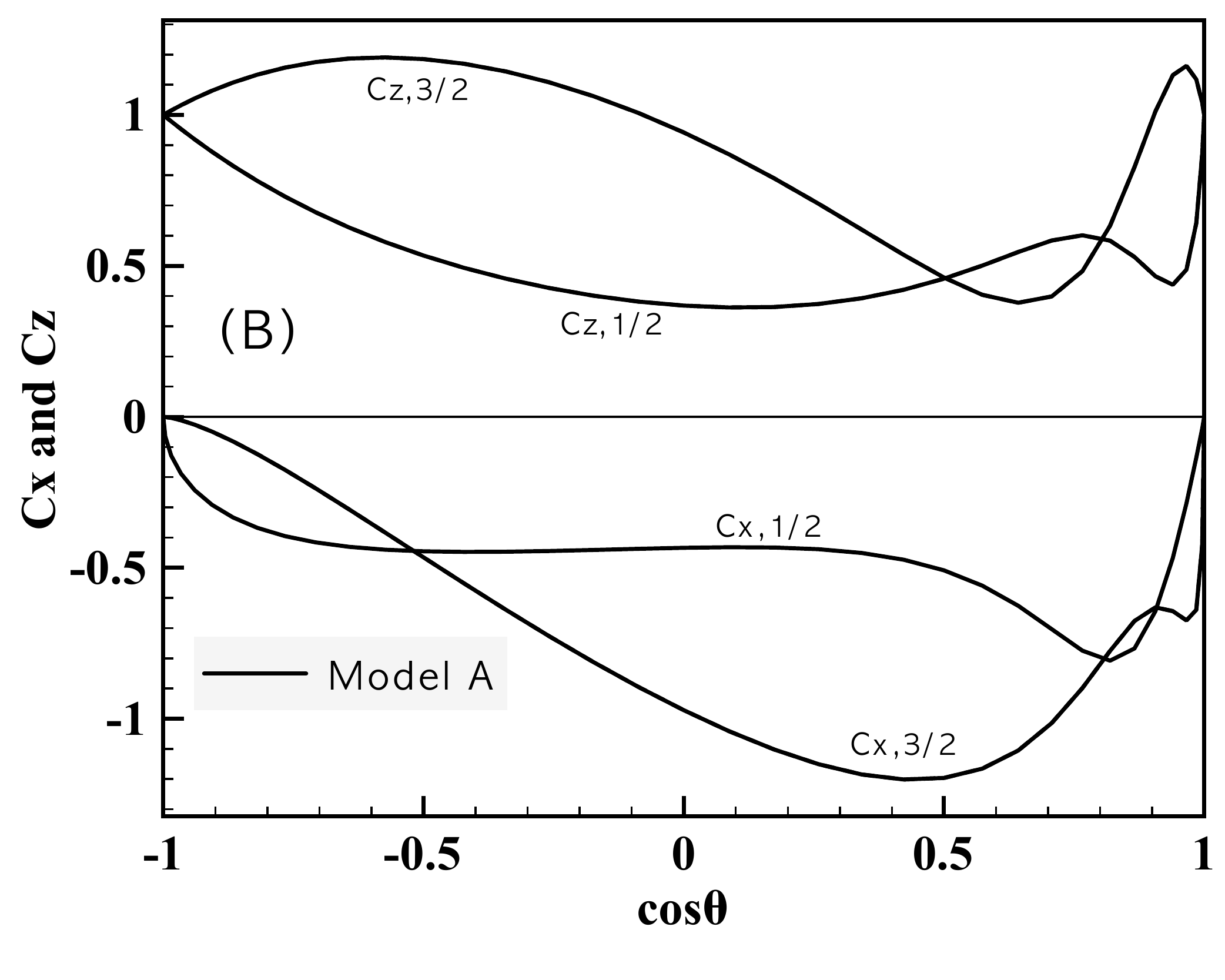}
\end{tabular}
\begin{tabular}{cc}
\includegraphics[width=7.5cm]{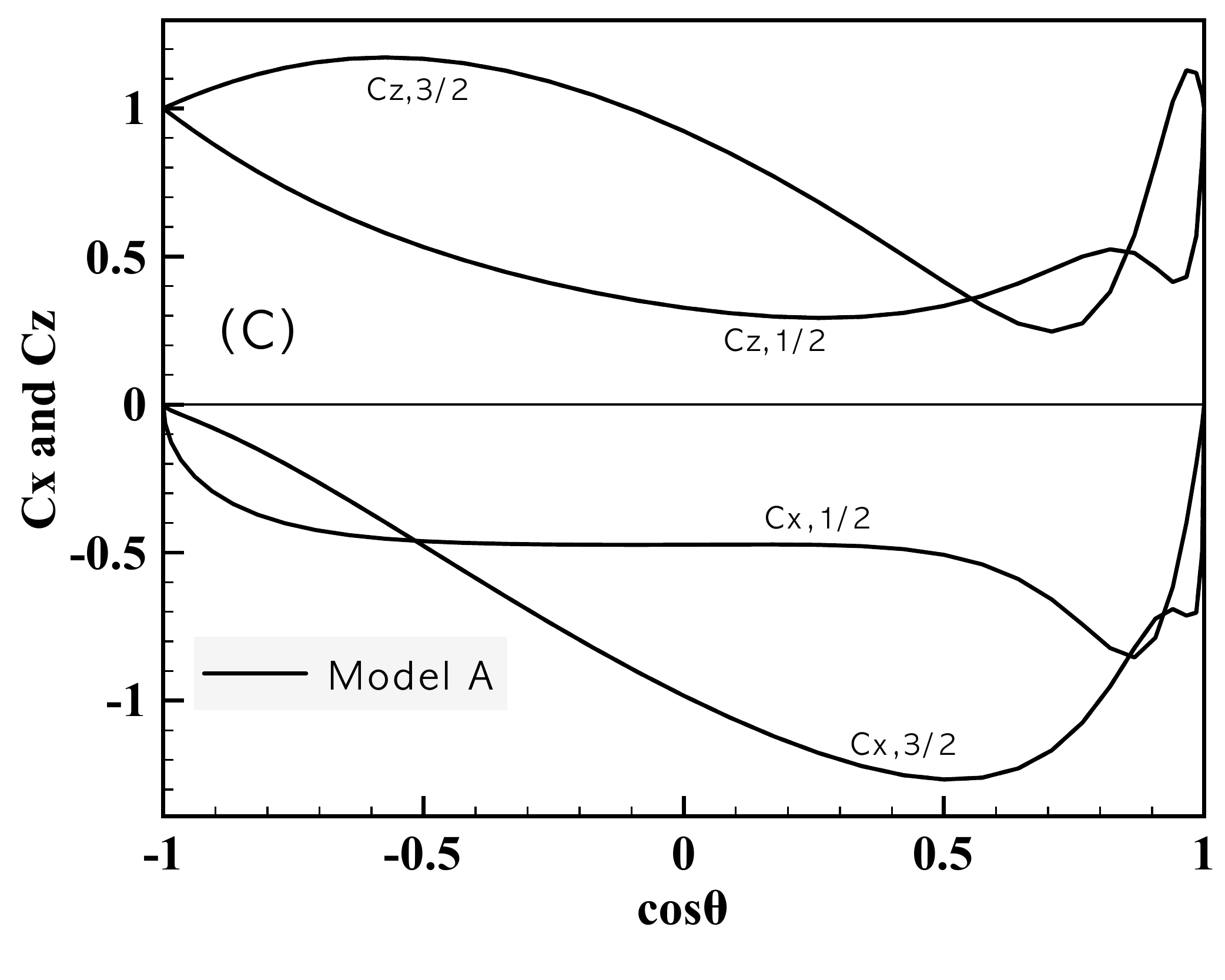}
\includegraphics[width=7.5cm]{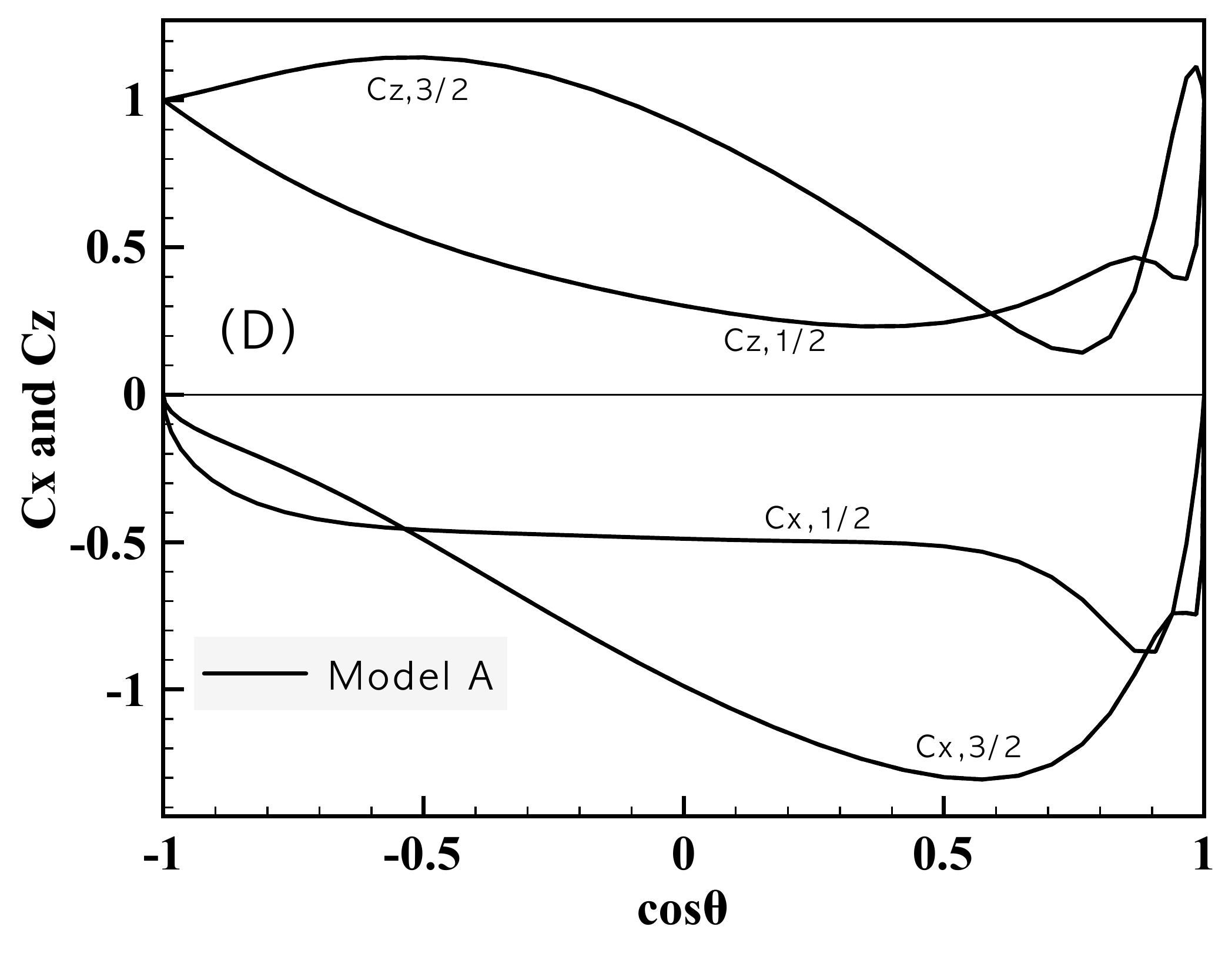}
\end{tabular}
\begin{tabular}{cc}
\includegraphics[width=7.5cm]{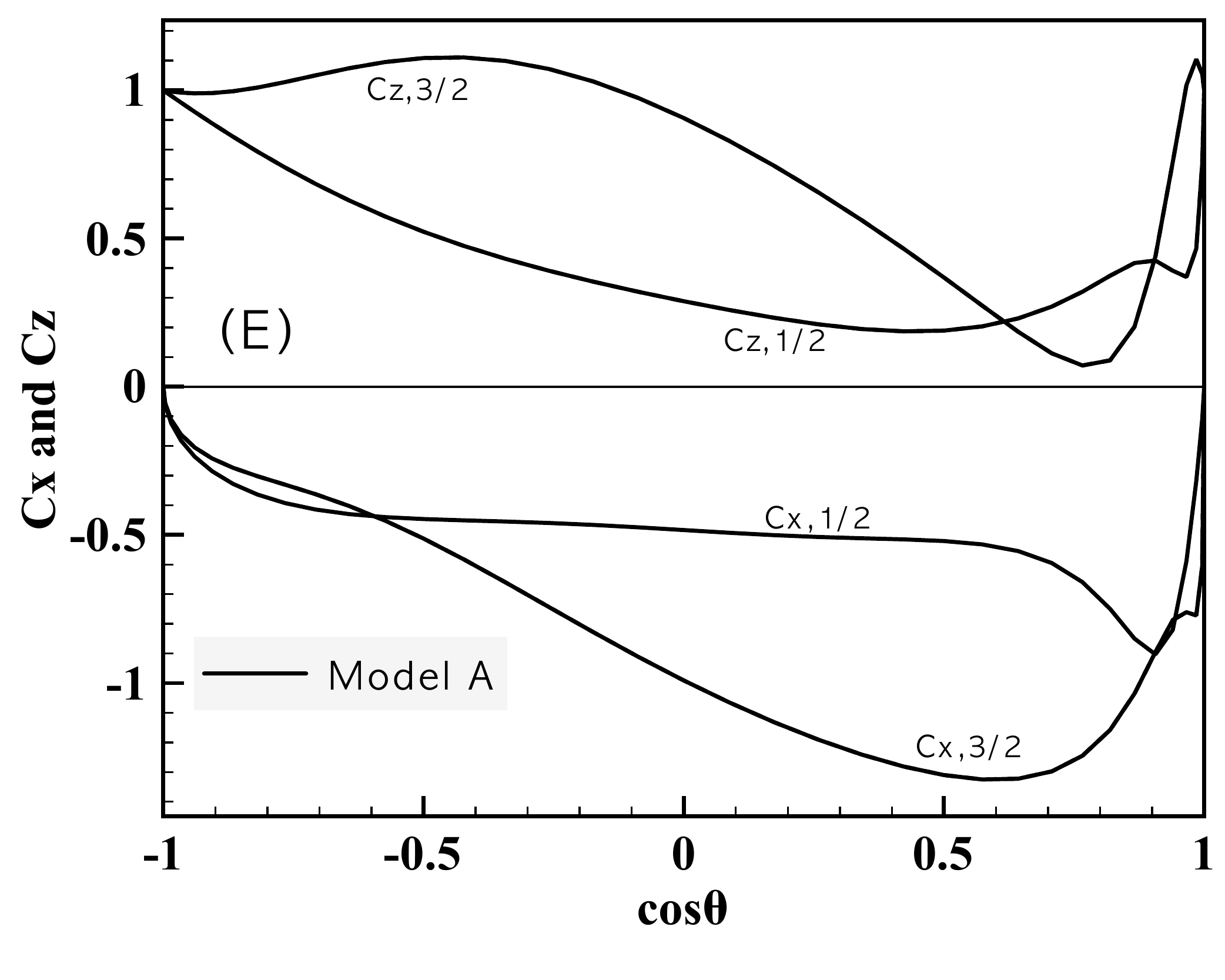}
\includegraphics[width=7.5cm]{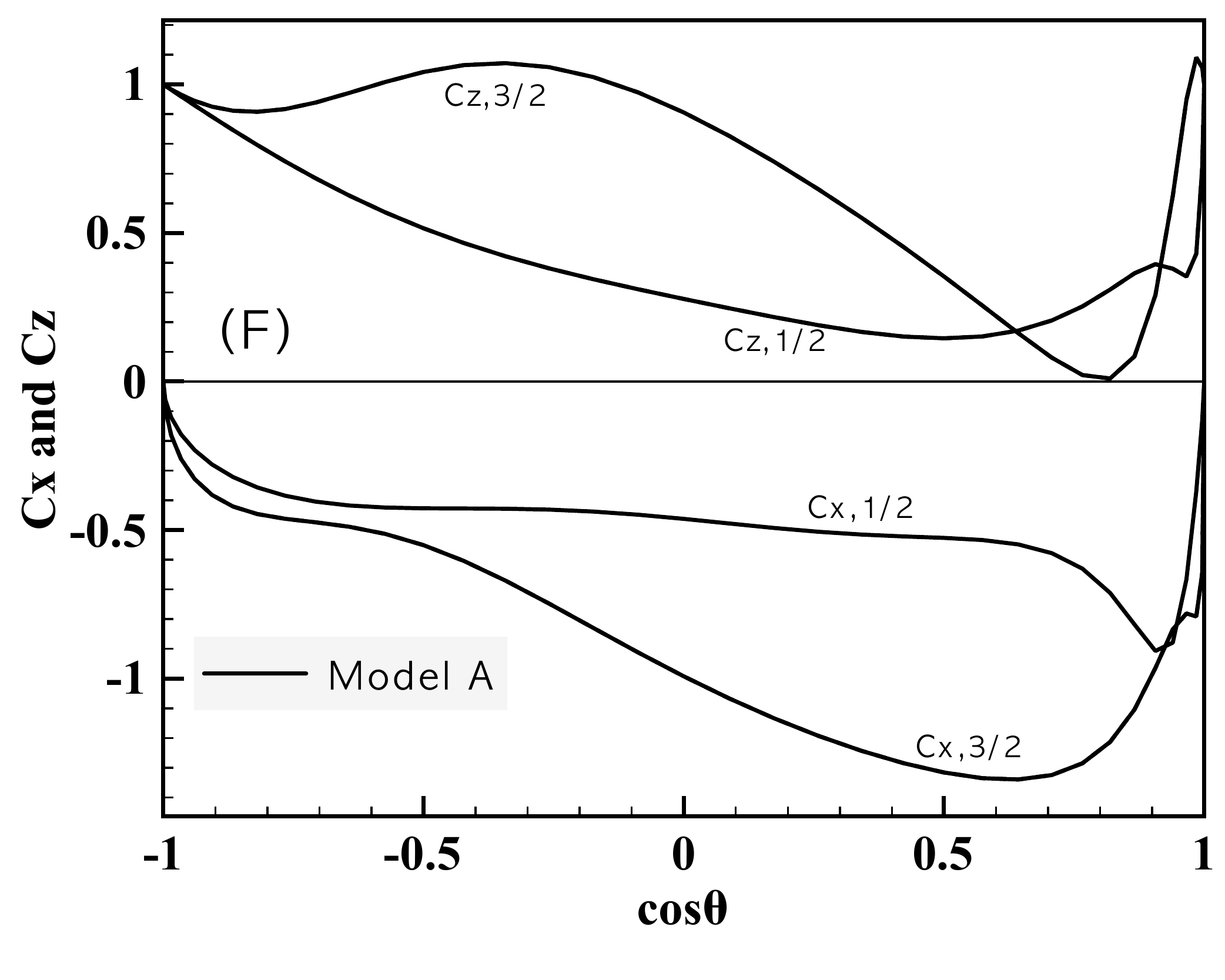}
\end{tabular}
\caption{Polarization-transfer coefficients, $C_{x,1/2}$, $C_{x,3/2}$, $C_{z,1/2}$, and $C_{z,3/2}$ as functions of $\cos\theta$ for $E_{\gamma}=2.4$ GeV (A), $2.9$ GeV (B), $3.4$ GeV (C), $3.9$ GeV (D), $4.4$ GeV (E), and $4.9$ GeV (F).Only the result of the model A are presented as explained in the text.}
\label{FIG10}
\end{figure}

In Figure~\ref{FIG10}, we depict the results of the polarization-transfer coefficients as functions of $\theta$ for $E_{\gamma}=2.4$ GeV (A), $2.9$ GeV (B), $3.4$ GeV (C), $3.9$ GeV (D), $4.4$ GeV (E), and $4.9$ GeV (F). Similarly to the photon-beam asymmetry, the Regge contributions are washed away again in these physical quantities, due to the same reason for the pohton-beam asymmetry understood by Eq.~(\ref{eq:CXZ}). Therefore the results of the model A and B show only negligible differences. It is quite different from the spin-$1/2$ $\Lambda(1116)$ photoproduction, in which a simple Regge model described experimental data qualitatively well~\cite{Bradford:2006ba}. The difference between the $\Lambda(1116)$ and $\Lambda^{*}$ photoproductions can be understood by the contact-term dominance in the later one. Moreover, the effects from resonances are of greater importance in the $\Lambda(1116)$ photoproduction~\cite{Janssen:2001pe}, than that of $\Lambda^{*}$~\cite{Nam:2009cv}.
Consequently, it is unlikely to have similar cancelation occurring in $\Lambda(1116)$ photoproduction
due to its complicated interference between the Born and resonance contributions.

\begin{figure}[ht]
\begin{tabular}{cc}
\includegraphics[width=7.5cm]{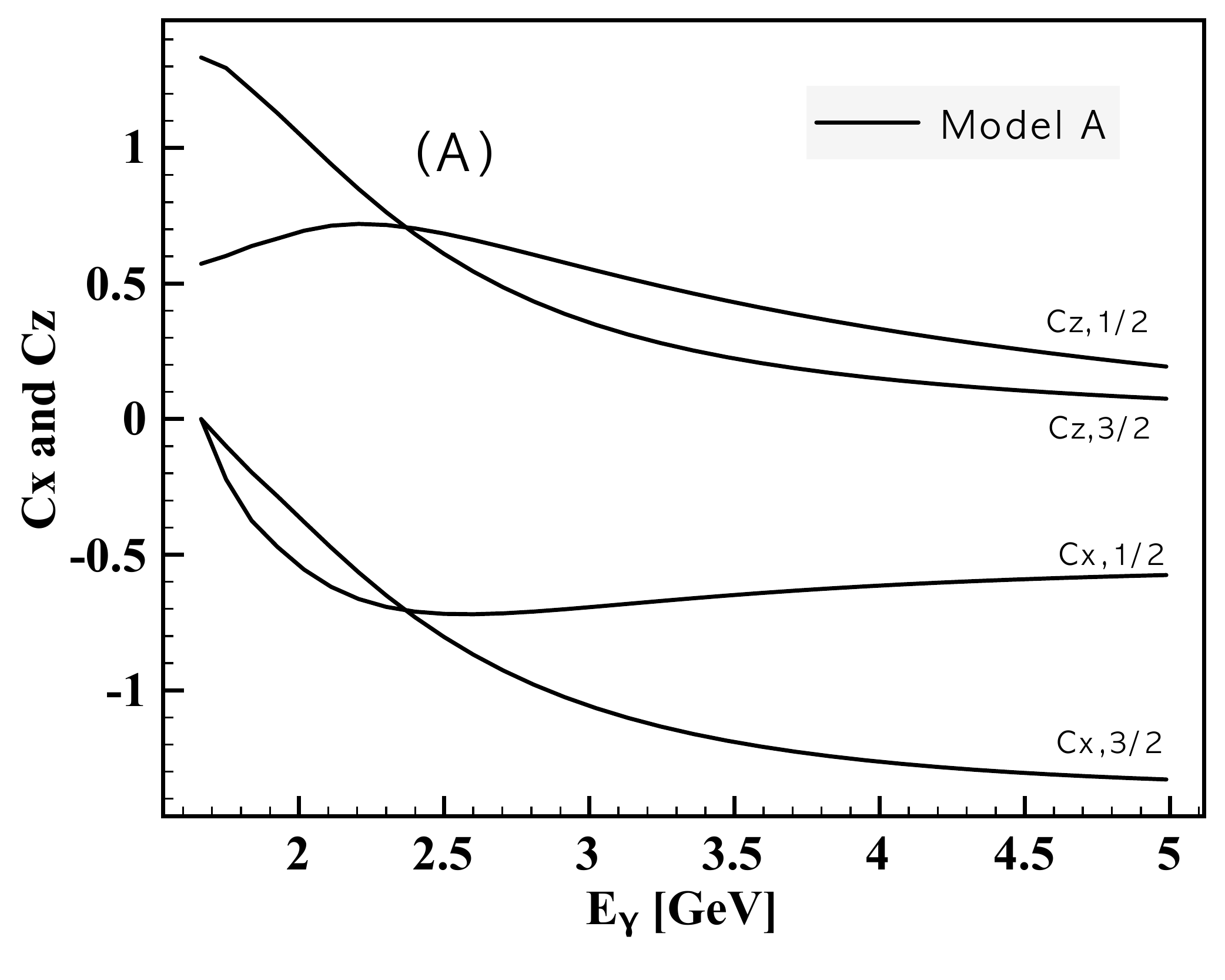}
\includegraphics[width=7.5cm]{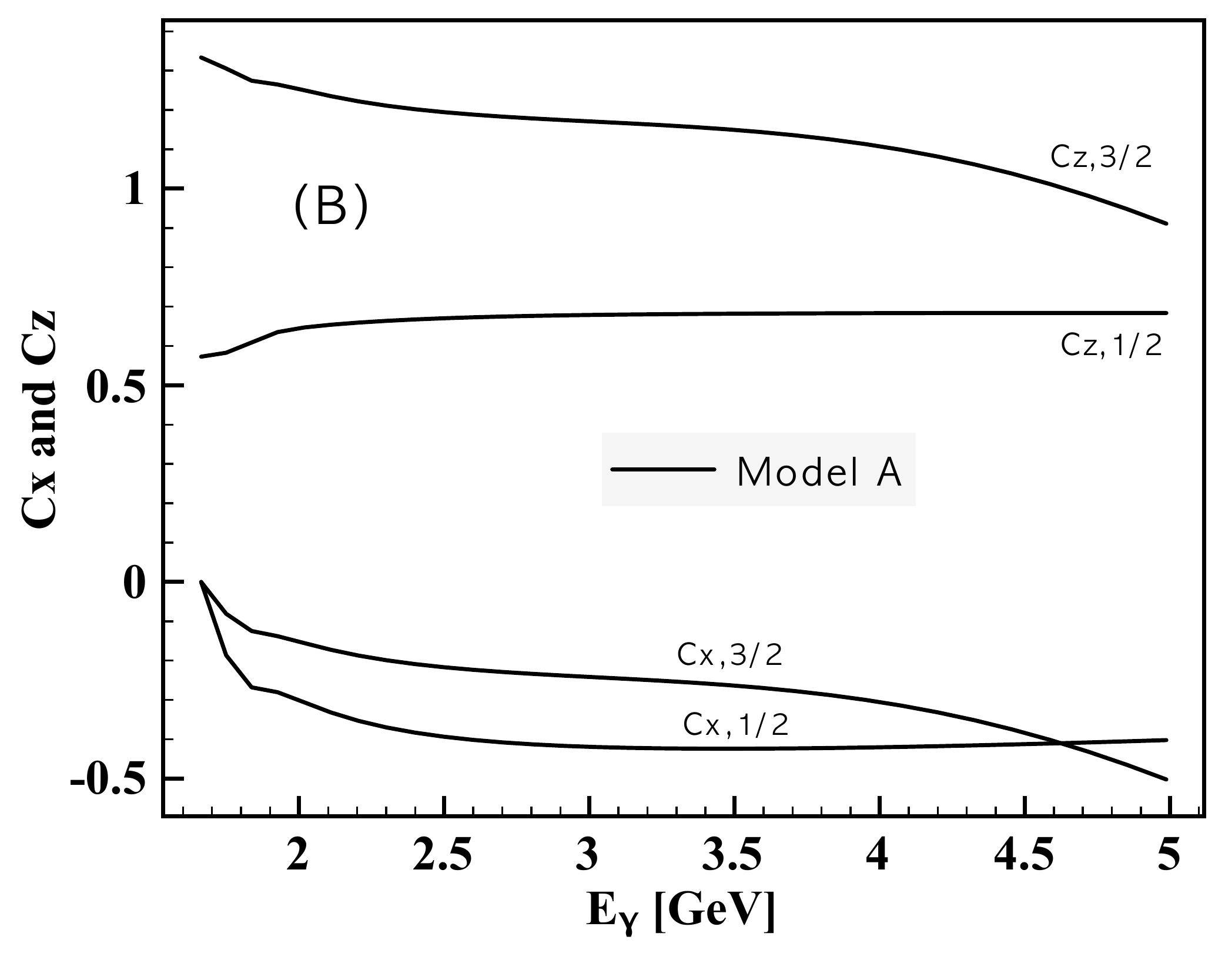}
\end{tabular}
\caption{Polarization-transfer coefficients, $C_{x,1/2}$, $C_{x,3/2}$, $C_{z,1/2}$, and $C_{z,3/2}$ as functions of $E_{\gamma}$ for $\theta=45^{\circ}$ (A) and $135^{\circ}$ (B). Only the result of the model A are presented as explained in the text.}
\label{FIG11}
\end{figure}

As discussed in~\cite{Nam:2009cv}, the shapes of the polarization-transfer coefficients are basically made of the contact-term contribution which provides symmetric and oscillating curves around zero and unity~\cite{Nam:2009cv}. The symmetric shapes are shifted into those shown in the Figure 10, because of the $\theta$-dependent $K$-exchange contribution providing complicated structures around $\cos\theta\approx0.5$. Interestingly, the results show that the shapes of the curves remain almost the same for all the the values of $E_{\gamma}$. Obviously visible differences start to appear for $E_{\gamma}\ge3.9$ GeV in the vicinity near $\cos\theta=-0.5$. We show the results of the polarization-transfer coefficients as functions of $E_{\gamma}$ in Figure~\ref{FIG11} for the two different angles, $\theta=45^{\circ}$ (A) and $180^{\circ}$ (B). Again the Regge contributions are negligible as expected.

\subsection{$K^{-}$-angle  distribution function}
Our last topic is the $K^{-}$-angle distribution function~\cite{Barber:1980zv,Muramatsu:2009zp} which is the angle distribution of $K^{-}$ decaying from $\Lambda^{*}$ ($\Lambda^{*}\to K^{-}p$) in the $t$-channel helicity frame, i.e. the Gottfried-Jackson frame~\cite{Schilling:1969um}. From this function, one can tell which meson-exchange contribution dominates the production process. According to the spin statistics, the function becomes $\sin^{2}\theta_{K^{-}}$ for $\Lambda^{*}$ in $S_{z}=\pm3/2$, whereas $\frac{1}{3}+\cos^{2}\theta_{K^{-}}$ for $\Lambda^{*}$ in $S_{z}=\pm1/2$. As in Ref.~\cite{Muramatsu:2009zp,Barrow:2001ds}, considering all the possible contributions, we can parametrize the function as,
\begin{equation}
\label{eq:DF}
\mathcal{F}_{K^{-}}
=A\sin^{2}\theta_{K^{-}}+B\left(\frac{1}{3}+\cos^{2}\theta_{K^{-}}\right),
\end{equation}
where $\mathcal{F}_{K^{-}}$ denotes the distribution function for convenience. The coefficients $A$ and $B$ stand for the strength of each spin state of $\Lambda^{*}$ with the normalization $A+B=1$. In principle there would be other hyperon contributions beside $\Lambda^{*}$ so that one can add an extra term to Eq.~(\ref{eq:DF}) representing the interference effects. However, we ignore this issue here for simplicity.

Before going further, it is worth mentioning about the experimental status for the quantity in hand. Note that each experiment provided a bit different result for $\mathcal{F}_{K^{-}}$. From LAMP2 collaboration~\cite{Barber:1980zv}, it was shown that $K^{-}$ decays mostly from $\Lambda^{*}$ in $S_{z}=\pm3/2$ state, showing a curve of $\mathcal{F}_{K^{-}}$ being close to $\sin^{2}\theta_{K^{-}}$ for $\theta=(20\sim40)^{\circ}$ ($A=0.880\pm0.076$ taken from Ref.~\cite{Muramatsu:2009zp}). On the contrary, using the data of electroproduction of $\Lambda^{*}$, CLAS collaboration showed rather complicated curves for $\mathcal{F}_{K^{-}}$ which is more or less close to that for the $S_{z}=\pm1/2$ state ($A=0.446\pm0.038$~\cite{Muramatsu:2009zp})~\cite{Barrow:2001ds}. Most recent experiment performed by LEPS collaboration, provided $\mathcal{F}_{K^{-}}$ for two different $\theta$-angle regions, $\theta=(0\sim80)^{\circ}$ and $\theta=(90\sim180)^{\circ}$. From their results, $\mathcal{F}_{K^{-}}$ looks similar to that for the $S_{z}=\pm3/2$ state in the backward region ($A=0.631\pm0.106$ via the side-band method~\cite{Muramatsu:2009zp}), whereas it shifts to the considerable mixture of the two states in the forward one ($A=0.520\pm0.063$~\cite{Muramatsu:2009zp}).

\begin{figure}[t]
\includegraphics[width=16cm]{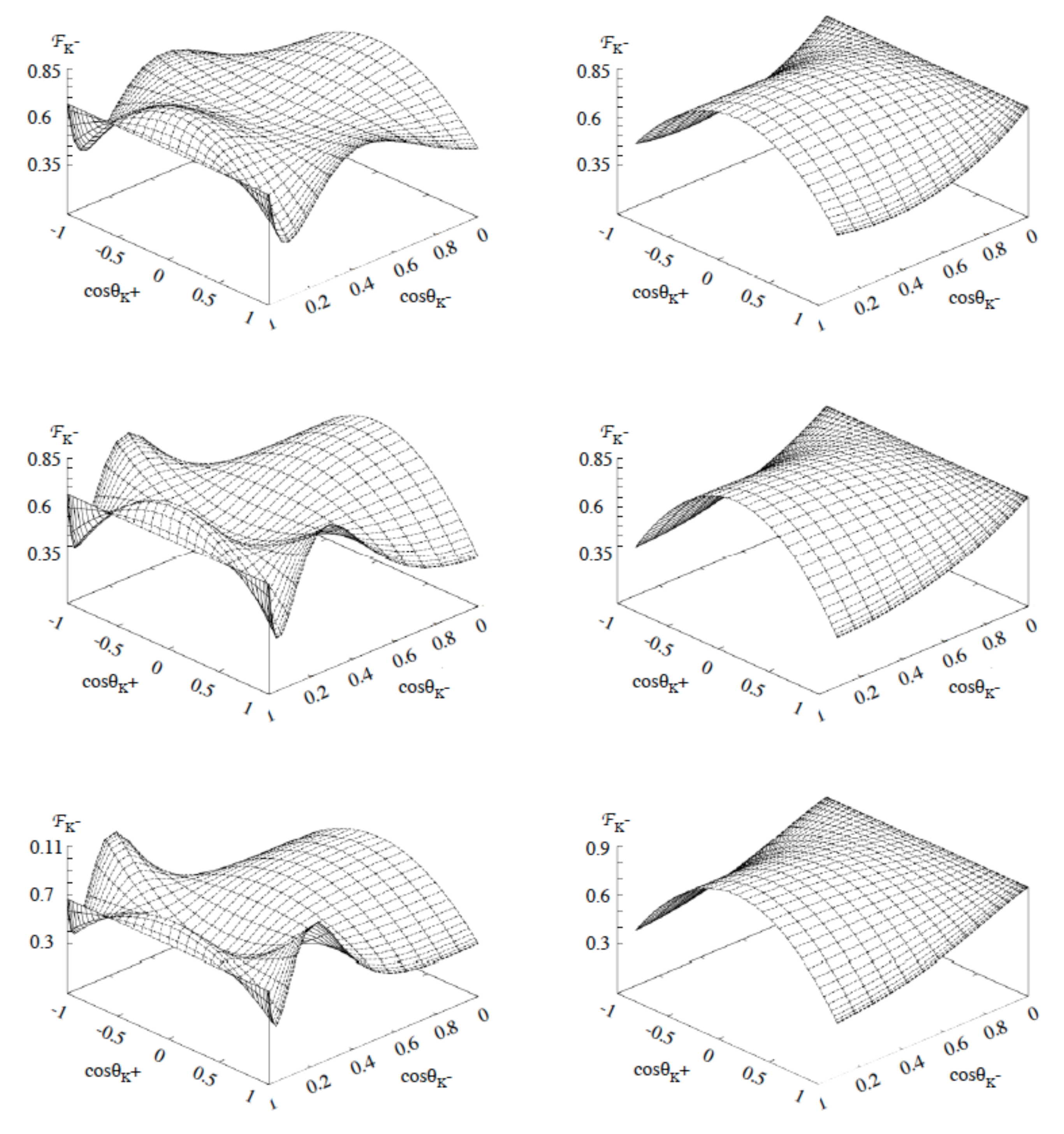}
\caption{$K^{-}$-angle distribution function $\mathcal{F}_{K^{-}}$ as a function of $\cos\theta_{K^{-}}$ and $\cos\theta_{K^{+}}$ for $E_{\gamma}\approx2.25$ GeV (first row), $3.25$ GeV (second row), and $4.25$ GeV (third row). The left column indicates the forward region, $\theta=(0\sim90)^{\circ}$, whereas the right column the backward region, $\theta=(90\sim180)^{\circ}$. Here, $\theta=\theta_{K^{+}}$.}
\label{FIG12}
\end{figure}

Here we want to provide our estimations on $\mathcal{F}_{K^{-}}$. Since the outgoing kaon ($K^{+}$) carries no spin, all the photon helicity is transferred to $\Lambda^{*}$ through the particle exchanged in the $t$-channel, Hence, it is natural to assume that the polarization-transfer coefficients in the $z$ direction should relate to the strength coefficients $A$ and $B$. Therefore, we express $A$ and $B$ in terms of $C_{z,1/2}$ and $C_{z,3/2}$ as follows:
\begin{equation}
\label{eq:AAA}
A=\frac{C_{z,3/2}}{C_{z,1/2}+C_{z,3/2}},\,\,\,\,
B=\frac{C_{z,1/2}}{C_{z,1/2}+C_{z,3/2}},
\end{equation}
In other words, $A$ denotes the strength that $\Lambda^{*}$ is in its $S_{z}=\pm3/2$ state, and $B$ for $S_{z}=\pm1/2$. In Figure~\ref{FIG12}, we depict $\mathcal{F}_{K^{-}}$ as a function of $\cos\theta$ and $\cos\theta_{K^{-}}$ at $E_{\gamma}=2.25$ GeV (first row), $2.35$ GeV (second row), and $4.25$ GeV (third row) for $\cos\theta=(0\sim1)$ (right column) and $\cos\theta=(0\sim-1)$ (left column). In the figure, we use the notation $\theta_{K^{+}}=\theta$. In general, we observe complicated mountains in the forward region, whereas the backward region shows simple sine curves ($\propto\sin^{2}\theta_{K^{-}}$ actually) for all the photon energies. In the vicinity near $\theta=0$, there is an area in which $\mathcal{F}_{K^{-}}\propto\sin^{2}\theta_{K^{-}}$. However, this area is shrunk as $E_{\gamma}$ increases. Just after this region, we face a second region where $\mathcal{F}_{K^{-}}\propto1+\cos^{2}\theta_{K^{-}}$. Again, this second region becomes narrower as $E_{\gamma}$ increases. After these regions and until $\theta\approx180^{\circ}$, $\mathcal{F}_{K^{-}}$ behaves as $\sin^{2}\theta_{K^{-}}$. From these observations, we conclude that the shape of $\mathcal{F}_{K^{-}}$ depends much on the value of $\cos\theta$ in the forward region, but insensitive to that in the backward one. In other words, unless we specify $\theta$ in the forward region, the shape of $\mathcal{F}_{K^{-}}$ can hardly be determined.

\begin{figure}[ht]
\begin{tabular}{cc}
\includegraphics[width=7.5cm]{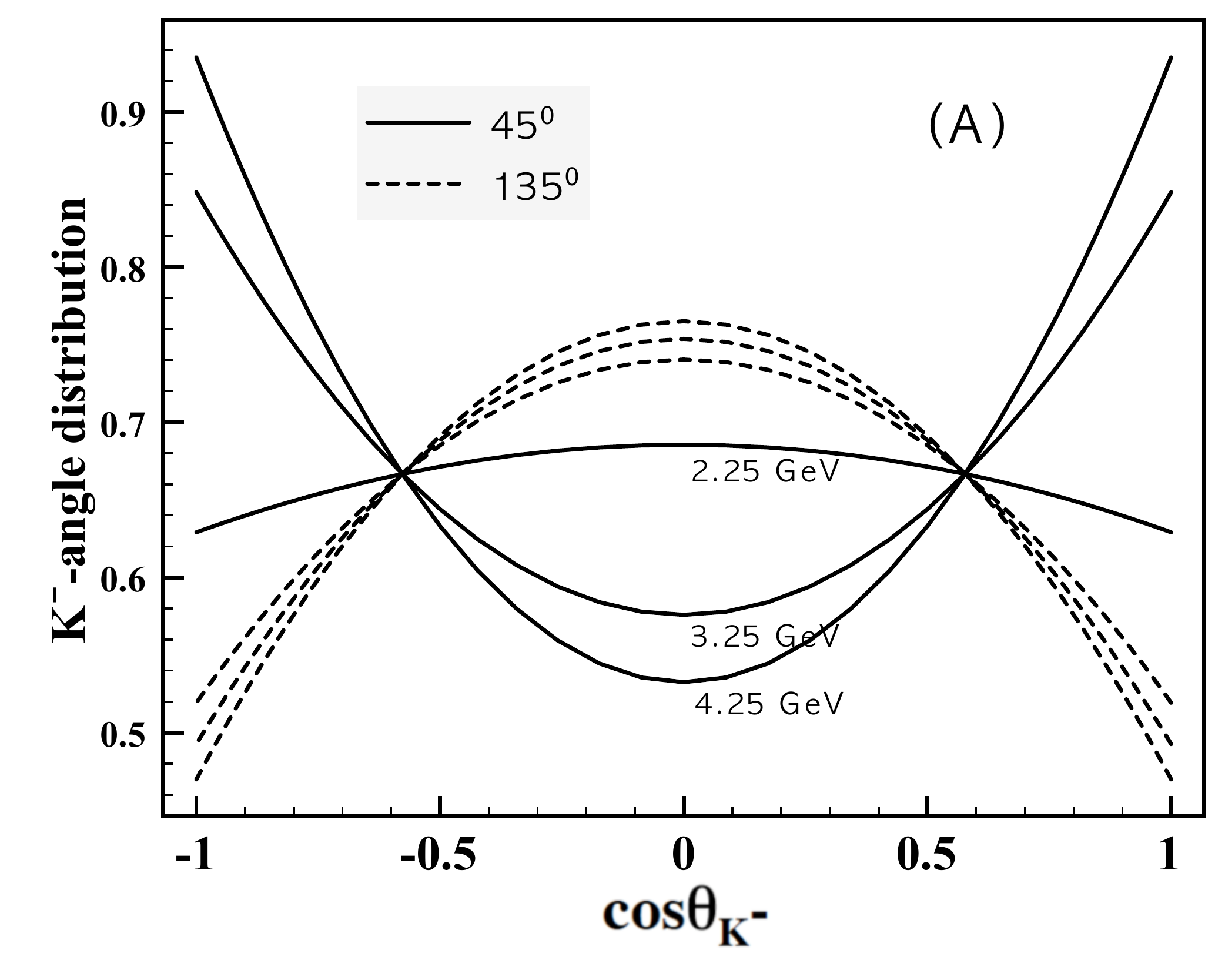}
\includegraphics[width=7.5cm]{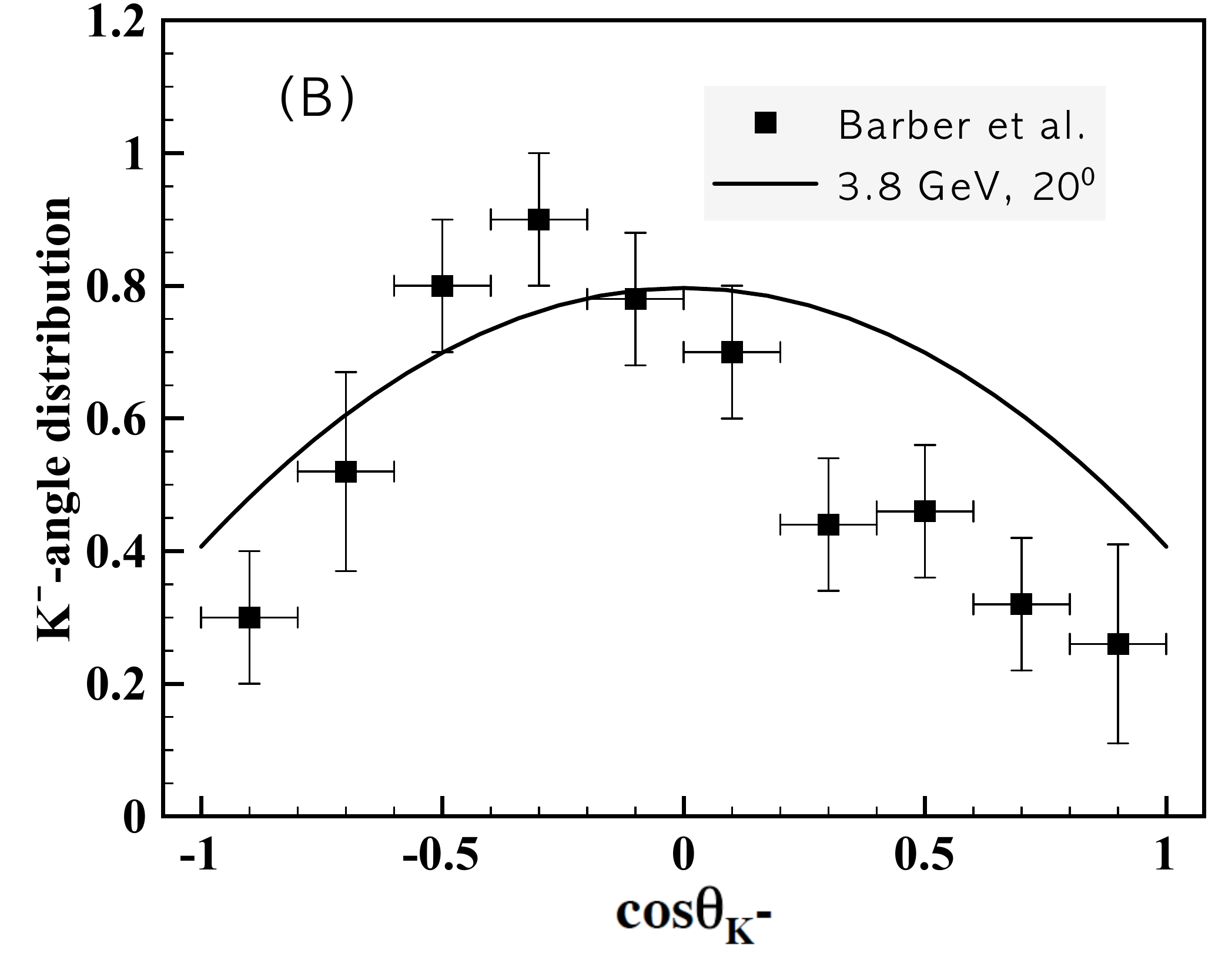}
\end{tabular}
\begin{tabular}{cc}
\includegraphics[width=7.5cm]{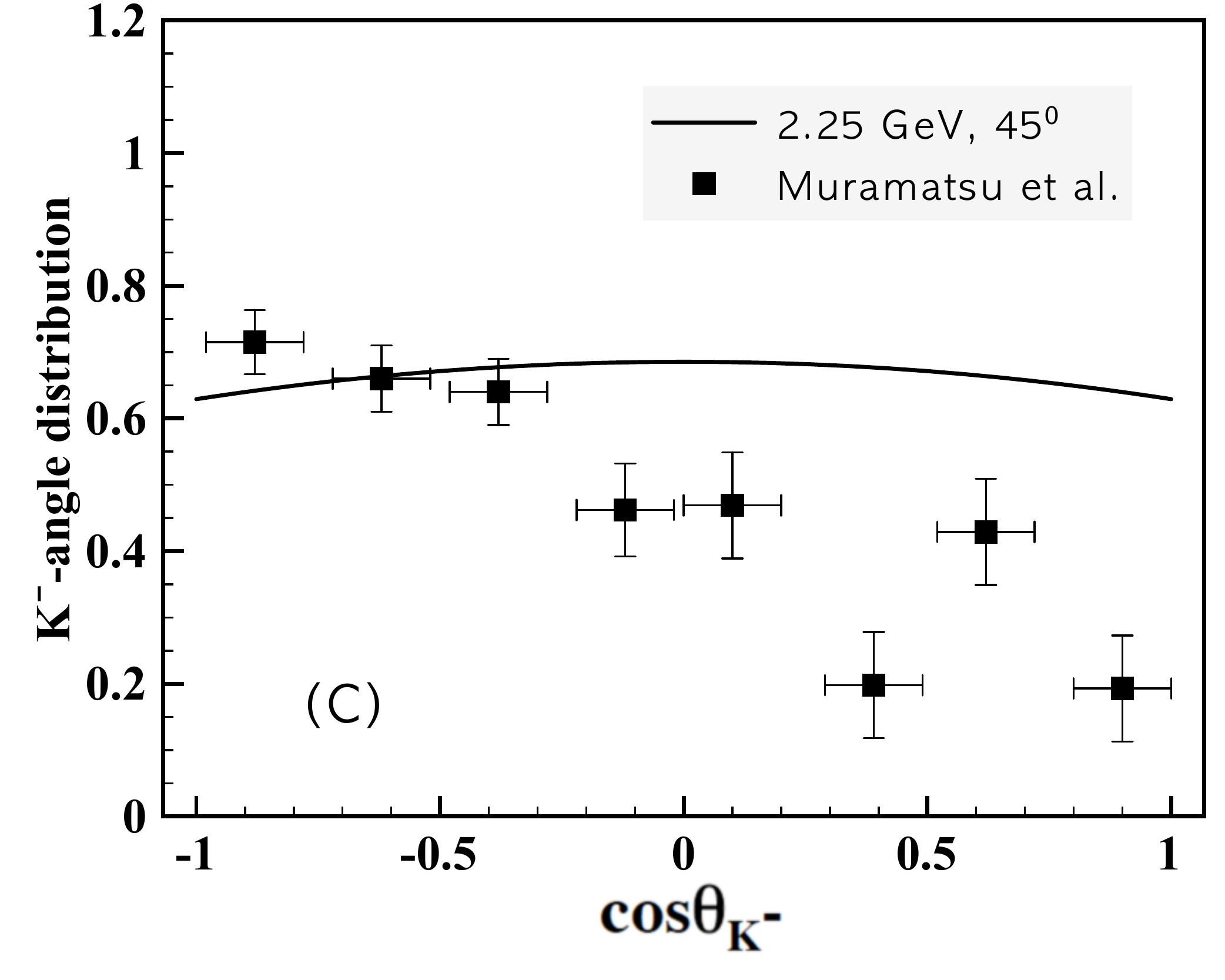}
\includegraphics[width=7.5cm]{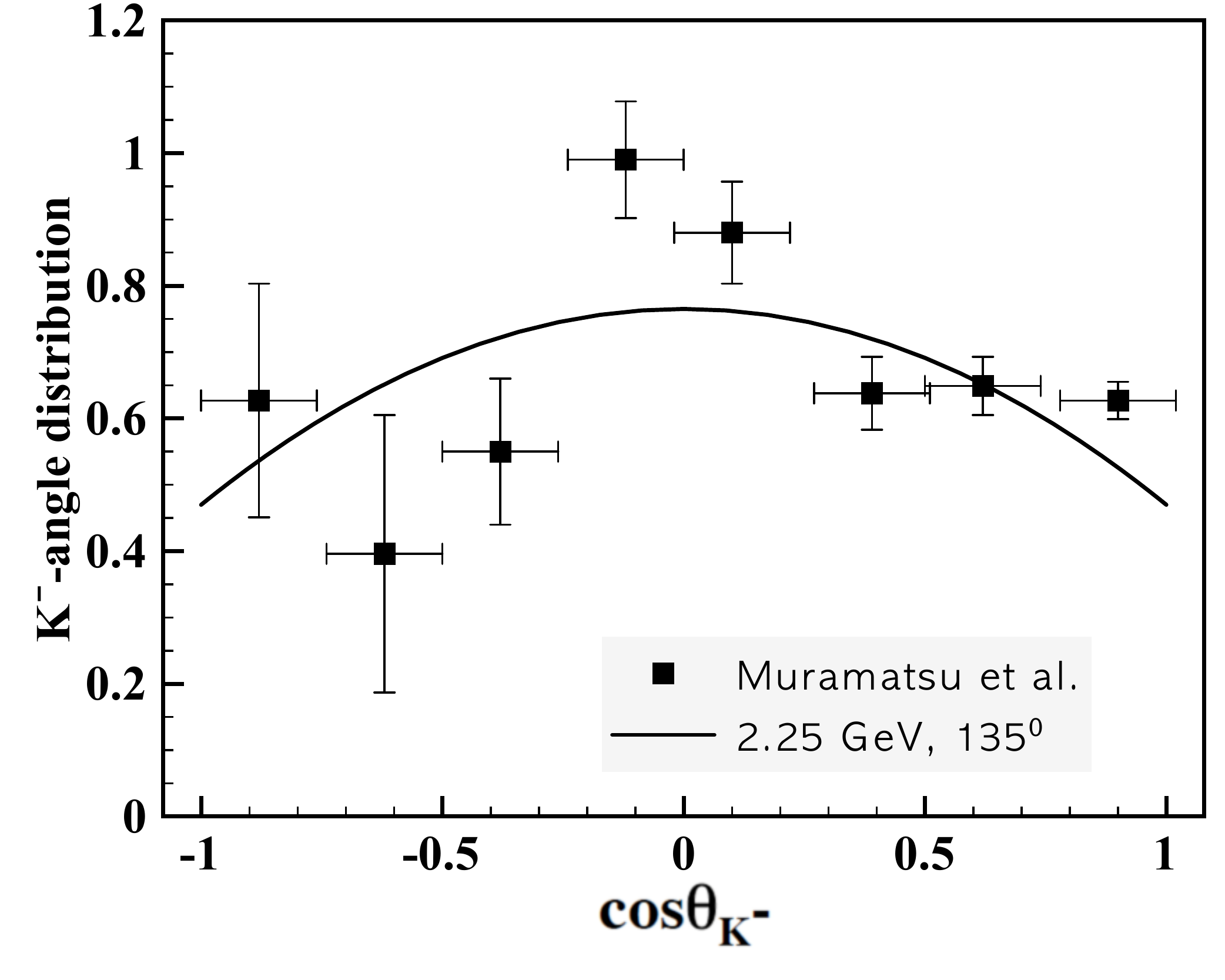}
\end{tabular}
\caption{$\mathcal{F}_{K^{-}}$ as a function of $\cos\theta_{K^{-}}$ for $E_{\gamma}=2.25$ GeV, $3.25$ GeV, and $4.25$ GeV at $\theta=45^{\circ}$ and $135^{\circ}$ in (A). In (B), we compare the numerical result for $E_{\gamma}=2.25$ GeV and $\theta=30^{\circ}$ with the experimental data taken from Ref.~\cite{Barber:1980zv}. Similarly, we show the comparisons in (C) and (D) for $\theta=45^{\circ}$ and $135^{\circ}$, respectively, with Ref.~\cite{Muramatsu:2009zp}. See the text for details.}
\label{FIG13}
\end{figure}

In (A) of Figure~\ref{FIG13}, $\mathcal{F}_{K^{-}}$ is plotted as a function of $\cos\theta_{K^{-}}$ for $E_{\gamma}=2.25$ GeV, $3.25$ GeV, and $4.25$ GeV at $\theta=45^{\circ}$ and $135^{\circ}$. In the backward region represented by $\theta=135^{\circ}$, the curves for $\mathcal{F}_{K^{-}}$ are similar to each other $\sim\sin^{2}\theta_{K^{-}}$, as expected from Figure~\ref{FIG12}. On the contrary, they are quite different in the forward region represented by $\theta=45^{\circ}$, depending on $E_{\gamma}$. This can be understood easily by seeing the left column of Figure~\ref{FIG12}; the curves, which are proportional to $\sin^{2}\theta_{K^{-}}$ or $\frac{1}{3}+\cos^{2}\theta_{K^{-}}$, are mixed, and the portion of each contribution depends on $E_{\gamma}$. In (B), we compare the numerical result for $E_{\gamma}=3.8$ GeV at $\theta=20^{\circ}$ with the experimental data taken from Ref.~\cite{Barber:1980zv} for $E_{\gamma}=(2.8\sim4.8)$ GeV and $\theta=(20\sim40)^{\circ}$. We normalize the experimental data with the numerical result by matching them at $\theta_{K^{-}}=90^{\circ}$ approximately. The theory and experiment are in a qualitative agreement, $\mathcal{F}_{K^{-}}\propto\sin^{2}\theta_{K^{-}}$.  Although we did not show explicitly, theoretical result for $\mathcal{F}_{K^{-}}$ changes drastically around  $\theta=25^{\circ}$. At $\theta\approx30^{\circ}$, the curve becomes $\sim\frac{1}{3}+\cos^{2}\theta_{K^{-}}$. This sudden change is consistent with the second row of Figure~\ref{FIG12}.

Similarly, we show the comparisons in (C) and (D) for $\theta=45^{\circ}$ and $135^{\circ}$, respectively, for $E_{\gamma}=2.25$ GeV with Ref.~\cite{Muramatsu:2009zp} for $E_{\gamma}=(1.75\sim2.4)$ GeV and $\theta=(0\sim180)^{\circ}$. Again, we normalized the experimental data to the numerical result for the backward-scattering region (D) as done above. Then, we used the same normalization for the forward-scattering region (C). As shown in (C), the experiment and theory start to deviate from each other beyond $\cos\theta_{K^{-}}\approx-0.2$. In Ref.~\cite{Muramatsu:2009zp}, it was argued that there can be a small destructive interference caused by the $K^{*}$-exchange contribution to explain the experimental data shown in (C). However, this is unlikely since that of $K^{*}$ exchange only gives negligible effect on $C_{z,1/2}$ and $C_{z,3/2}$~\cite{Nam:2009cv}. Hence, we consider the large deviation in (C) may come from the interference between $\Lambda^{*}$ and other hyperon contributions which are not taken into account in the  present work. As in the backward region, $\mathcal{F}_{K^{-}}$ shows a curve $\sim\sin^{2}\theta_{K^{-}}$, and the experimental data behaves similarly. We list numerical values of $A$ calculated using Eq.~(\ref{eq:DF}), in Table~\ref{TAB3} for  $\theta=45^{\circ}$ and $135^{\circ}$. Although we have not considered the interference, for these specific angles, present theoretical  estimations on $A$ are very similar to those given in Ref.~\cite{Muramatsu:2009zp} as seen in the table.

\begin{table}[b]
\begin{tabular}{c||c|c|c||c|c|c}
&$2.25$ GeV&$3.25$ GeV&$4.25$ GeV
&\cite{Muramatsu:2009zp}
&\cite{Barber:1980zv}
&\cite{Barrow:2001ds}\\
\hline
\hline
$45^{\circ}$&$0.528$ &$0.364$ &$0.299$&$0.520\pm0.063$&$0.880\pm0.076$&$0.446\pm0.038$\\
\hline
$135^{\circ}$&$0.648$ &$0.631$ &$0.611$&$0.631\pm0.106$&-&-
\end{tabular}
\caption{Coefficient $A$ in Eq.~(\ref{eq:DF}). The row and column represent $\theta$ and $E_{\gamma}$, respectively. The values for $A$ for Refs.~\cite{Muramatsu:2009zp,Barber:1980zv,Barrow:2001ds} are taken from Ref.~\cite{Muramatsu:2009zp}.}
\label{TAB3}
\end{table}

\section{Discussion and Summary}
In this article, we have investigated the $\Lambda^{*}$ photoproduction off the proton target within a hadronic model including
the tree-level diagrams with the nucleon and the certain resonance intermediate states
and the Regge contributions. We computed the energy and angular dependences of the cross section and the polarization observables in the production process. We employed the gauge-invariant form factor scheme developed in Refs.~\cite{Haberzettl:1998eq,Davidson:2001rk,Haberzettl:2006bn}. Taking into account the fact that the Regge contributions become important at $(s,|t|)\to(\infty,0)$ and remain non negligible even for $(s,|t|)\to(s_{\mathrm{threshold}},\mathrm{finite})$, we adopt an interpolating ansatz to incorporate the physical situations.
With the inclusion of the Regge contributions, we followed the prescription of Refs.~\cite{Corthals:2005ce,Corthals:2006nz,Ozaki:2009wp} to preserve the gauge invariance. The common form factor $F_{c}$ is also replaced by the one modified by the Regge contributions. The important observations in the present work are summarized as follows:
\begin{itemize}
\item All the physical observables computed are comparable with the current experimental data except the data of the energy dependence at very backward direction.
\item The Regge contributions are necessary to explain the experimental data such as the angular dependences in the high-energy region.
\item From our results, the Regge contributions become significant beyond $E_{\gamma}\gtrsim4$ GeV in the forward region, $|t|\lesssim0.2\,\mathrm{GeV}^{2}$.
\item Bump structures appear in the angular dependences of the cross section in the forward regions above $E_{\gamma}\approx4$ GeV, which is due to the Regge contributions.
\item The $K^{*}$-exchange contribution can be ignored rather safely. The $D_{13}$ contribution turns out to be very small with the parameters extracted from current data and the quark model calculation~\cite{Capstick:1998uh}.

\item The polarization observables are insensitive to the Regge contributions because the cancelation of the relevant form factors. This can be understood by the contact-term dominance as a consequence of the gauge invariance.
\item  The photon-beam asymmetry $\Sigma$ is largely dominated by the $K$-exchange contribution. The polarization-transfer coefficients is determined by the contact term contribution and the $\theta$-dependent $K$-exchange effect.
\item The $K^{-}$-angle distribution function, $\mathcal{F}_{K^{-}}$ shows the mixture of the curves proportional to $\sin^{2}\theta_{K^{-}}$ or $\frac{1}{3}+\cos^{2}\theta_{K^{-}}$ for the forward $K^{+}$ scattering region. In the backward region, it remains almost unchanged over $\theta$, showing the curve proportional to $\sin^{2}\theta_{K^{-}}$, indicating the spin-$3/2$ state of $\Lambda^{*}$ manifestly.
\end{itemize}

It was reported that LEPS and CLAS collaborations have planned to
upgrade their photon energies up to $E_{\gamma}\approx3$ GeV (LEPS2)
and $12$ GeV (CLAS12 especially for GPD physics), respectively.
Although the LEPS upgrade energy is not enough to see the Regge
contributions which starts to be effective over $E_{\gamma}\approx5$
GeV, it is still desirable because the low-energy data is important
to test whether the Regge contributions are absent or not up to that
energy, comparing with the present theoretical results. Moreover,
their linear and circular polarization data will be
welcomed~\cite{Kohri}. Since the angular and polarization
observables show significantly different behaviors for the Regge
contributions, the measurements of these quantities beyond
$E_{\gamma}\approx4$ GeV must be a crucial test of our model whose
essence is the contact-term dominance in terms of gauge invariance.
In addition, the theoretical estimations on the coefficient $A$ of
$\mathcal{F}_{K^{-}}$ will be a good guide to analyze the
experiments.

As mentioned already, there are still rooms to accommodate unknown $s$- and $u$-channel contributions which may improve the agreement between present model and the experimental data to a certain extent. In particular, the $u$-channel physics may play an important role in reproducing the rise in the backward region. Related works are underway and will appear elsewhere.

\section*{acknowledgment}
The authors are grateful to A.~Hosaka, T.~Nakano, H.~-Ch.~Kim, H.~Kohri, and S.~Ozaki for fruitful discussions. S.i.N. appreciates the technical support from Y.~Kwon. S.i.N was supported by the grant NSC 98-2811-M-033-008 and
C.W.K  was supported by the grant NSC 96-2112-M-033-003-MY3 from National Science Council (NSC) of Taiwan.
The support from National Center for Theoretical Sciences (North) of Taiwan (under the grant number NSC 97-2119-M-002-001) is also acknowledged. The numerical calculations were performed using MIHO at RCNP, Osaka University, Japan. 

\newpage
\end{document}